\newcommand{\OG}{ \texttt{OpenGadget3 }}
\newcommand{\quantities}[1]{%
  \begin{tabular}{@{}c@{}}\strut#1\strut\end{tabular}%
}

\documentclass{aa}  

\usepackage{graphicx}
 \usepackage{hyperref}
\usepackage{txfonts}
\usepackage{xcolor}
\usepackage{booktabs} 
\usepackage{subcaption}
\usepackage{enumitem}

\usepackage{soul}

\usepackage{color}

\definecolor{boh}{rgb}{1,0,0}

\setlist{leftmargin=5.5mm}
\begin{document}

\title{Dynamical friction and the evolution of black holes in cosmological simulations: A new implementation in \OG}

   \author{Alice Damiano\inst{1,2,3,4,5},
          Milena Valentini \inst{1,2,4,5},
          Stefano Borgani\inst{1,2,3,4,5},
          Luca Tornatore\inst{2,5},
          Giuseppe Murante \inst{2,3,5},
          Antonio Ragagnin \inst{3,6,7},
          Cinthia Ragone-Figueroa \inst{8,2},
          Klaus Dolag \inst{9,10}
          }

   \institute{Dipartimento di Fisica dell'Universit\`a di Trieste, Sez. di Astronomia, via Tiepolo 11, I-34131 Trieste, Italy    
     \and
      INAF -- Osservatorio Astronomico di Trieste, via Tiepolo 11, I-34131, Trieste, Italy 
        \and
        IFPU, Institute for Fundamental Physics of the Universe, Via Beirut 2, 34014 Trieste, Italy
        \and
        INFN, Instituto Nazionale di Fisica Nucleare, Via Valerio 2, I-34127, Trieste, Italy
        \and
        ICSC - Italian Research Center on High Performance Computing, Big Data and Quantum Computing, via Magnanelli 2, 40033, Casalecchio di Reno, Italy
        \and
        INAF-Osservatorio di Astrofisica e Scienza dello Spazio di Bologna, Via Piero Gobetti 93/3, I-40129 Bologna, Italy
        \and
        Dipartimento di Fisica e Astronomia "Augusto Righi", Alma Mater Studiorum Università di Bologna, via Gobetti 93/2, I-40129 Bologna, Italy
         \and
        Instituto de Astronom\'ia Te\'orica y Experimental (IATE), Consejo Nacional de Investigaciones Cient\'ificas y T\'ecnicas de la\\ Rep\'ublica Argentina (CONICET), Universidad Nacional de C\'ordoba, Laprida 854, X5000BGR, C\'ordoba, Argentina
         \and
        Universit\"ats-Sternwarte M\"unchen, Scheinerstr. 1, D-81679, M\"unchen, Germany
        \and
         Max-Plank-Institut f\"ur Astrophysik, Karl-Schwarzschild Strasse 1, D-85740 Garching, Germany
        }

  \abstract{}
  {We introduce a novel sub-resolution prescription to correct for the unresolved dynamical friction (DF) onto black holes (BHs) in cosmological simulations, to describe BH dynamics accurately, and to overcome spurious motions induced by numerical effects.}  
  {We implemented a sub-resolution prescription for the unresolved DF onto BHs in the \OG code. We carried out cosmological simulations of a volume of $(16 \ \mathrm{comoving~Mpc})^3$ and zoomed-in simulations of a galaxy group and of a galaxy cluster. We assessed the advantages of our new technique in comparison to commonly adopted methods for hampering spurious BH displacements, namely repositioning onto a local minimum of the gravitational potential and ad hoc boosting of the BH particle dynamical mass. We inspected variations in BH demography in terms of offset from the centres of the host sub-halos, the wandering population of BHs, BH--BH merger rates, and the occupation fraction of sub-halos. We also analysed the impact of the different prescriptions on individual BH interaction events  in detail.}
  {The newly introduced DF correction enhances the centring of BHs on host halos, the effects of which are at least comparable with those of alternative techniques. Also, the correction becomes gradually more effective as the redshift decreases. Simulations with this correction predict half as many merger events with respect to the repositioning prescription, with the advantage of being less prone to leaving substructures without any central BH. Simulations featuring our DF prescription produce a smaller (by up to $\mathrm{\sim 50 \%}$ with respect to repositioning) population of wandering BHs and final BH masses that are in good agreement with observations. Regarding individual BH--BH interactions, our DF model captures the gradual inspiraling of orbits before the merger occurs.  By contrast, the repositioning scheme, in its most classical renditions, describes extremely fast mergers, while the dynamical mass misrepresents the dynamics of the black holes, introducing numerical scattering between the orbiting BHs.}
 { The novel DF correction improves the accuracy if tracking BHs within their hosts galaxies and the pathway to BH-BH mergers. This opens up new possibilities for better modeling the evolution of BH populations in cosmological simulations across different times and different environments.}

\titlerunning{Dynamical friction onto BHs in cosmological simulations}

\authorrunning{Damiano A. et al.}
   \maketitle

%

\section{Introduction}
Supermassive black holes (SMBHs) reside at the centres of massive galaxies and are considered to affect their evolution profoundly. Numerous studies provided evidence of the relation between the mass of a  SMBH and the properties of its host galaxy \citep[e.g.][]{kormendy1993nearest, magorrian_demography_1998, ferrarese_fundamental_2000, gebhardt_relationship_2000, merritt_mbh-sigma_2000, ferrarese_supermassive_2001, haering_black_2004, gultekin_m-sigma_2009, graham_log-quadratic_2007, mcconnell_revisiting_2013, gaspari_x-ray_2019}. The most widely accepted explanation is that during SMBH growth by gas accretion, a small fraction of the enormous amount of the released gravitational energy couples with the surrounding environment, regulating the galaxy star formation via various possible mechanisms \citep[e.g.][]{silk_quasars_1998,  granato.etal.2004, hopkins_unified_2005,
Bower_2006,
cattaneo_role_2009, gitti_evidence_2012}.

Given the influence the SMBHs play in shaping the environment where they reside, it is essential to trace their dynamics correctly. Massive BHs are thought to be formed at some early epochs through mechanisms ranging from direct collapse of primodial gas clouds or as the end stage of very massive Population III stars \citep[e.g.][]{bromm2003, Volonteri_2012, Mayer_2018}, and to subsequently grow in the dense cores of galaxies. Nonetheless, recent studies have consistently shown cases of SMBHs exhibiting substantial displacements from their host galaxies \citep[e.g.][]{Webb_2012,Menezes_2014, Combes_2019, Reines_2020}. The dynamical behaviour of SMBHs is significantly affected by the dynamical friction (DF) force \citep{chandrasekhar, binneytremaine} exerted by the matter distribution surrounding them. This drag force in general prevents an SMBH from escaping the centre of its host galaxy, is responsible for the migration of BH to the galaxy centre, and drives the initial stages of a merger event between two SMBHs, ultimately leading to the formation of a close pair \citep{begelmann}. 

Cosmological simulations represent ideal tools for following the  evolution of structures given the highly nonlinear astrophysical phenomena governing the interactions of baryonic matter ---including gas and stars--- and also interactions with dark matter (DM).
N-body simulations describe the gravitational instability of a collisionless fluid, which is sampled by a discrete set of `macro particles' \citep[e.g.][for reviews]{Borgani_2011,Springel2016}. For this reason, tracking the orbits of single collisionless particles has little physical relevance, as it is their ensemble properties that carry the most importance.
A BH particle, on the other hand, 
is introduced in cosmological simulations as an individual collisionless particle, and has a specific physical meaning; its presence is capable of significantly impacting the global properties of the galaxy it belongs to.  
Unlike the surrounding macro-particles, BHs are not entities whose motion can be interpreted solely on a global scale. Instead, their motion mirrors the effective motion of an astrophysical object. 
The contrast between the nature of BH particles in these simulations and the interpretation of the surrounding ones presents a primary conceptual obstacle when tracking the trajectories of BH particles. 

Specifically, scattering interactions between a BH and its surrounding particles can `heat' the BH orbit.
A spurious, numerical displacement of a BH from the centre of the host galaxy is a major consequence of this heating, and can eventually lead to the formation of unwanted `wandering' BHs. Furthermore,  this displacement also negatively impacts the capability of simulations to describe BH--BH merger events, ultimately leading to an incorrect description of AGN feedback, which in turn strongly affects the predictions of the simulations. For instance, \cite{Ragone_Figueroa_2018} found that a better centring of the SMBH within the host galaxy in cluster simulations was key to predicting the masses of the brightest cluster galaxies (BCGs; confirmed by comparison with observations), which were overpredicted by a factor of a few in their previous work \citep{Ragone_2013}.

The issues described above stem from numerical simulations failing to recover the DF force.
As the magnitude of the drag due to DF  depends on the interactions between BHs and the surrounding particles, any limitation in reconstructing the correct gravitational interactions at the N-body resolution level results in an inaccurate representation of this effect.
In this context, the first question that we attempt to answer here pertains to whether it is feasible to introduce a correction to the gravitational acceleration provided by the N-body solver that accounts for the unresolved DF. 

Instead of relying on some numerical `tricks' to control the dynamics of the BHs, such as artificially repositioning the BHs at the position of a local minimum of the gravitational potential  \citep{Springel_2005, Di_Matteo_2008, Sijacki_2015, dave_simba_2019, Ragone_Figueroa_2018, Bassini_2019, Bahe2022}, or using a boosted dynamical mass to enhance the effect of the resolved DF \citep{DeBuhr_2011, Bassini_2020}, some authors have already suggested addressing this problem by introducing an explicit correction for the unresolved DF \citep{hirschmann_cosmological_2014, tremmel_off_2015, astrid, Chen, ma_new_2023}. 
For instance, \cite{hirschmann_cosmological_2014} proposed the application of a DF correction given by the Chandrasekhar DF formula. In their application, the maximum impact parameter entering in the DF correction is the  half-mass radius of the substructure hosting the BH. In addition, \cite{tremmel_off_2015} stated that, under the hypotheses of a sufficiently shallow potential surrounding the BH, only unresolved interactions within the softening length require correction. Still, \cite{Chen} proposed that the application of the DF correction should be coupled with a boosted dynamical mass in order to account for those cases in which a BH has a rather small mass, close to its value at seeding, and is located in a poorly resolved halo.
The absence of a consensus on the possibility, use, and actual computation of a DF correction to improve the description of the dynamics of BHs in cosmological simulations is what motivated this study.

In this paper, we propose a novel implementation of the DF correction, which we realise using the \OG code for cosmological N-body and hydrodynamical simulations. 
In this implementation, a correction to DF force is computed by explicitly accounting for the contributions of numerical particles whose gravitational interactions with a BH particle ---as provided by the N-body solution--- are directly affected by force softening. 
As we extensively discuss in the following, the primary advantage of our approach is that it is less affected by the assumptions on which the derivation of the Chandrasekhar DF formula is based.
In more detail, we aim to address the following questions: Does our approach provide an adequate description of the DF force acting on BHs? How does it compare against the other numerical ah hoc prescriptions (i.e. repositioning and dynamical mass) introduced to mimic the effect of DF on BH particles? 
To answer these questions, we simulate a group-sized halo and a cluster-sized halo, with initial conditions from the \texttt{DIANOGA} set \citep{Bonafede_2011, Bassini_2019, Bassini_2020}, along with a cosmological box with a comoving size of 16 Mpc per side (16 cMpc). 
The simulations were carried out three times, maintaining identical settings for all parameters, except for the sub-resolution prescription governing the BH dynamics: continuous repositioning on a local potential minimum, boosted dynamical mass, and our novel model to correct the unresolved DF. 
This approach enables us not only to focus on systems densely populated and rich in interactions, such as galaxy clusters and groups, but also to carry out a statistical analysis of the BH population in a cosmological volume.

The paper is organised as follows.
In Sect. \ref{Dynamics} we present the new DF model and compare it with both previous DF corrections and numerical prescriptions to constrain the BH dynamics.
In Sect. \ref{GeneralBHs} we describe the implementation of SMBH evolution and the ensuing AGN feedback in the\OG code. After introducing the details of our test simulations in Sect. \ref{Simulations}, we present the results of our analysis of the general properties of the SMBH population  in Sect. \ref{GlobalProperties}. In Section \ref{zoom} we zoom in, limiting our analysis to single BHs and merger episodes, and compare the small-scale effects of our sub-resolution model of DF.

\section{Dynamics of BH particles} \label{Dynamics}

Due to the finite mass and force resolution of cosmological simulations, the effect of DF exerted on a BH particle by surrounding particles is always underestimated and, in general, subject to discreteness noise. Such limitations can lead to a grossly incorrect description of the orbits of BH particles, which leads in turn to an incorrect description of the ensuing AGN feedback and of the predictions on the SMBH population. We refer to Sect. \ref{GeneralBHs} for details on the BH seeding, accretion and feedback mechanisms in cosmological hydrodynamical simulations. In this section we present different approaches, including our new one, which have been introduced to overcome this limitation. In Sect. \ref{newmodel}, we present our new description of a correction to the gravitational acceleration to account for the unresolved DF. 
Previous works already proposed approaches to correct DF for finite-resolution effects, and some details of our implementation revisit their arguments. For this reason, we highlight in Sect. \ref{others} the conceptual differences between our method and such previous approaches proposed in the literature. Finally, Sect. \ref{pinning} and Sect. \ref{dynamicalmass} describe other methods, based on ad hoc prescriptions, to correct BH dynamics for the unresolved DF.
\subsection{A new prescription for dynamical friction} \label{newmodel}
To clearly describe in detail the DF correction proposed in this work, we review the basic steps of
 the DF expression originally derived by \cite{chandrasekhar}, which holds under the assumption of specific hypotheses. 
The starting point is a system of two particles moving on a Keplerian orbit around each other. The velocity variation $\Delta \vec{v}_{\rm M}$ of a particle of mass $\mathrm{M}$ caused by the interaction with another particle of mass $m$ and velocity $\vec{v}$ can be expressed in terms of their impact parameter $b$ and relative velocity, $\vec{v}_0=\vec{v}-\vec{v}_{\rm M}$, when the two particles were initially at large distance.  The components of $\Delta \vec{v}_{\rm M}$ parallel and perpendicular to the direction of $\vec{v}_0$ can be expressed as \citep{binneytremaine}:
\begin{equation} \label{v_parallel}
    |\Delta \vec{v}_{\rm M}|_{\parallel} = \frac{2m|\vec{v}_0|}{({M}+m)}\left[ 1+\frac{b^2|\vec{v}_0|^4}{G^2({\rm M}+m)^2}\right]^{-1}\,;
\end{equation}
\begin{equation} \label{v_perp}
    |\Delta \vec{v}_{\rm M}|_{\perp} = \frac{2mb|\vec{v}_0|^3}{G({\rm M}+m)^2}\left[ 1+\frac{b^2|\vec{v}_0|^4}{G({\rm M}+m)^2}\right]^{-1}\,.
\end{equation}
If the particle $M$ moves in a ‘sea' of particles of mass $m$, then the DF force acting on the former arises as the sum of the contributions of the velocity variations given by eqs. \eqref{v_parallel} and \eqref{v_perp}, due to the interactions with all the surrounding particles. In the derivation by \cite{binneytremaine}, the mass $m$ of the ‘sea' particles is assumed to be the same for all such particles. 
Under the assumption that the distribution of particles is uniform around the BH, the perpendicular contributions of Eq. \eqref{v_perp} sum to zero. 
The rate of encounters having impact parameter in the range $[b,b+db]$ is then given by: $2 \pi \,b \, db \ {{| \Delta \vec{v}_{M}|}_{\parallel}} f(\vec{v}) d^3 \vec{v}$, where $f(\vec{v})$ is the phase-space number density of stars.
Integrating this rate over the impact parameter from $0$ to $b_{ \rm max}$ \footnote{In principle, this integration should be performed from a minimum value of the impact parameter equal to the Schwarzschild radius of the BH. It is unnecessary to consider this contribution to derive an expression for our DF correction. Therefore, we assume the minimum value of $b$ to be zero.}, we have that the DF force acting on a BH of mass M from particles having mass $m$ and velocities in the range $(\vec{v},  \vec{v}+d^3\vec{v})$ is:
\begin{equation} \label{start}
\frac{d \vec{v}_{ \rm M}}{dt} \bigg|_{\vec{v}
} = 2 \pi \ln\left[{1+{\Lambda 
(m, \vec{v})}^2}\right]G^2 m ({\rm M}+m) f(\vec{v}
) d^3 \vec{v} \frac{(\vec{v}-\vec{v}_{ \rm M})}{|\vec{v}-\vec{v}_{\rm M}|^3}\, ,
\end{equation}
where
\begin{equation}
\Lambda (m, \vec{v})=\frac{b_{\rm max}(\vec{v}-\vec{v}_{ \rm M})^2}{G({\rm M}+m)}.
\end{equation}
Here, $b_{\rm max}$ is the maximum impact parameter. In general, $b_{\rm max}$is interpreted as the largest distance from the target BH particle, which contains all the particles contributing to the DF exerted on the BH itself. In the above expression, we have made explicit the dependence of $\Lambda$ on mass $m$ and velocity $\vec{v}$.

The maximum allowed value of the impact parameter should in principle be set by the size of the system containing all the particles contributing to the DF exerted on a BH particle. In fact, different assumptions for the values of $b_{\rm max}$ have been introduced in the literature, all having a degree of arbitrariness. In the next section, we provide an extended comparison between different choices. As in \cite{tremmel_off_2015}, we assume $b_{\rm max}$ to be given by the gravitational softening length of the BH particle, $\epsilon_{\rm BH}$, meaning that above such length, the N-body solver is assumed to provide already a correct description of DF. 
In the context of our simulations, the BH is surrounded by particles which trace the underlying continuous density field. Each of these particles has its mass, $m_i$, and its velocity $\vec{v}_{m,i}$. 
The phase-space number density of particles surrounding the BH can be then expressed as a sum of delta-functions, each with a mass $m_i$ and a velocity $\vec v_ {{m},i}$:
 \begin{equation} \label{velocitydistrib}
 f(\vec{v}) = \frac{3}{4 \pi {\epsilon_{\rm BH}}^3} \sum_{i=1}^{N(<\epsilon_{\rm BH})} \delta (\vec{v}-\vec v_{m,i})\,,
 \end{equation}
where the sum is over all the $N(<\epsilon_{\rm BH})$ particles lying at a distance from the BH smaller than its gravitational softening scale. 

Equation \eqref{velocitydistrib} relies on the hypothesis that the particles within the softening length provide an adequate sampling of the velocity field of the ‘sea' surrounding the BHs. To validate this assumption, we performed several tests presented in Appendix \ref{appendix_distrbution}.
For well-resolved sub-halos, where the number of particles within the softening length can be as large as $\sim 10^{4}$, this assumption is valid.

In the simulations performed in this work, we only use DM and star particles to compute the correction to DF, while we defer to a forthcoming analysis the inclusion of gas particles (see for example \citealt{Dubois_2014}).
Using Eqs. \eqref{start} and \eqref{velocitydistrib}, we can then derive the total DF force term, by integrating over the surrounding particles' velocities:
 \begin{equation} \label{new}
\frac{d \vec{v}_{\rm M}}{dt}  =   \frac{3G^2}{2 {\epsilon_{\rm BH}}^3} \sum_{i=1}^{N(<\epsilon_{\rm BH})} \ln\left[{1+{\Lambda} (m_i)^2}\right]  m_i ({\rm M}+m_i) \frac{(\vec{v}_{m,i}-\vec{v}_{\rm M})}{|\vec{v}_{m,i}-\vec{v}_{ \rm M}|^3}\,.
\end{equation}
 
The gravitational acceleration of the BH is then corrected by the DF acceleration  $\vec{a}_{\rm df}$ given by Eq. \eqref{new}, so that the total acceleration acting on a single BH particle is:
\begin{equation}
\vec{a}_{\rm BH}=\vec{a}_{\rm g}+\vec{a}_{\rm df}\,,
\end{equation}
where $\vec{a}_{\rm g}$ is the acceleration provided by the N-body solver. It is important to remark that the DF correction hence obtained is a contribution correcting for the softened interactions between the BH and the surrounding particles, but not for the absence of  information on the sub-softening structure of phase-space.

In order to validate the performance of this model in a controlled numerical experiment, without the complexity of a full-physics simulation in a cosmological context, we carry out several tests for the sinking timescale of a BH initially placed on a circular orbit and infalling toward the centre of an isolated DM halo, and compare them to analytical predictions. The results of these tests are reported in Appendix \ref{appensixsinking}. They show in general that simulations which include our correction for unresolved DF are in very good agreement with analytical predictions, and the convergence with resolution to this analytical result is faster than when DF correction is neglected.

We also point out that our approach for the computation of the DF correction presents some advantages with respect to the models proposed in the literature. We discuss this in the next section, providing an overview of the implementations adopted nowadays and underlining their differences with our approach. 

\subsection{Previous approaches} \label{others}
Correcting the unresolved DF force acting on BH particles using a physically motivated approach instead of resorting to ad hoc prescriptions is clearly highly desirable.
 This is the reason why, the implementation of a correction of the DF force acting on BH particles, based on the original derivation by \cite{chandrasekhar}, has been explored in various studies since the very first implementations (e.g. \citealt{hirschmann_cosmological_2014}, \citealt{tremmel_off_2015}).
A common aspect of all such approaches is that they start from Eq. \eqref{start}, and implicitly assume that the BH is surrounded by a homogeneous and infinite distribution of particles all having the same mass. However, different approaches make different assumptions of the physical size of the region where to correct for the unresolved DF, i.e. on the actual value of $b_{\rm max}$. For instance, in the work by \cite{hirschmann_cosmological_2014}, $b_{\rm max}$ is defined as the typical size of the system hosting the BH and, as such, it is set to the half-mass radius of the sub-halo\footnote{This parametrisation required an on-the-fly execution of the {\tt{SubFind}} algorithm \citep{Springel.etal.2001, subfind} to identify the substructures hosting the BHs and to compute their half-mass radius.}. On the other hand, \cite{tremmel_off_2015} argued that DF is correctly computed by the N-body solver at scales larger than the gravitational softening length of the BH, $\epsilon_{\rm BH}$. Accordingly, a correction term to the DF should be added only on scales smaller than $\epsilon_{\rm BH}$, so that $b_{\rm max}=
\epsilon_{\rm BH}$. 
In line with this approach, \cite{Pfister_2019} accounted for the DF force from particles within a sphere centred on the BH position and with radius which is a multiple of the adaptive grid mesh size on which the gravitational force is computed \citep{teyssier2002}. Finally, the DF model recently presented by \cite{Chen} and \cite{astrid} assumes a constant value of $b_{max}=10 h^{-1}$ckpc and 20 kpc, respectively. 
As for the velocity distribution function of the sea of particles around the BH, \cite{hirschmann_cosmological_2014}, \cite{Chen} and \cite{astrid} adopt the same standard hypothesis, originally formulated by \cite{binneytremaine}, of a local Maxwellian velocity distribution. Under the further hypothesis that $M_{\rm BH} \gg m$, where $m$ is the mass of the particles around the BH, the DF force $\vec{F}_{\rm DF}$ can be cast in the form
\begin{equation}\label{eq:DF_BT}
\vec{F}_{\rm DF} = -4\pi\rho \left(\frac{G{\rm M_{ \rm BH}}}{{v}_{\rm BH}} \right)^2 \mathrm{F}(x) \ln(\Lambda) \widehat{v}_{\rm BH}\,.
\end{equation}
Here,  $\rho$ is the smoothed density at the position of the BH, contributed by stellar and DM particles, using the BH smoothing length. Furthermore, $\widehat{v}_{\rm BH}$ is the versor of the BH velocity relative to the ‘sea' of surrounding particles, while
\begin{equation}
    \mathrm{F}(x)={\rm erf}(x) -\frac{2x}{\sqrt{\pi}}e^{-x^2} \ \ ; \ \ x = \frac{v_{\rm BH}}{\sigma_v},
\end{equation}
with $\sigma_v$ the velocity dispersion of the surrounding particles. 
We assume that stars and DM particles exert a DF force on the BHs. 
\cite{Ostriker_1999} computed the contribution to DF from gas, lately included in simulations as an additional numerical corrective term by \cite{Chen}. However, in their analysis, \cite{Chen} found that the DF correction is in fact dominated by the collisionless component.

Rather than assuming a specific expression for the velocity distribution function of the particles around a BH, \cite{tremmel_off_2015} proposed to incorporate the mass $m$ of the surrounding particles within the integral over velocities, thus moving the uncertainty on the velocity distribution function to an uncertainty on the mass density of the surrounding particles. 
The approach by \cite{tremmel_off_2015} has been then applied also in \cite{Bellovary_2018}. They found that the effect of DF correction is efficient for BHs having a mass at least three times larger than that of the surrounding particles. This result justifies the choice made by \cite{Chen} to include both a DF correction and a boosted dynamical mass for BH particles (see Sect. \ref{dynamicalmass} below). 

Finally, a scheme that stands out from the others is the one proposed by \cite{ma_new_2023}. In their approach, a discrete N-body correction, similar to the one proposed in this paper, is taken into account, but still acting only on scales above the gravitational force resolution of simulations. 
Differently from such an approach, the model that we described in Sect. \ref{newmodel} explicitly intends to correct for the  interactions that take place below the BH softening scale which, by definition, are not correctly described by the N-body solver. 
Overall, the main differences between the DF correction proposed in this work and the previous ones are the absence of any a priori assumption on the velocity distribution of the surrounding particles and the relaxation of the hypothesis that the surrounding particles' mass is negligible compared to the BH one. Our approach accounts for single scatters between the BH and each particle within the BH softening, each contributing with its own velocity and mass. In this way, we try to reduce the dynamical heating of BHs as a consequence of the noisy background potential, taking place whenever their mass is comparable to or even lower than the surrounding particles' one. In summary, Table \ref{tab: differences} lists the main differences between our novel DF correction introduced in Sect.\ref{newmodel} and those proposed by \cite{hirschmann_cosmological_2014}, \cite{tremmel_off_2015}, and \cite{Chen}.

\begin{table*}
        \centering
        \caption{Main differences between the DF proposed in this work and others.}
        \label{tab: differences}
        \begin{tabular}{lccccr}
                \hline
        \toprule 
                 & (1) &  (2) & (3) \\
        \hline
                \midrule
                 \quantities{This \\ work} & \quantities{Relaxation of the hypothesis \\ $ \rm m \ll M_{BH}$ \\ \\ 
   $\rm b_{max} = \epsilon_{BH}$ instead of the  \\ half-mass radius of the host sub-halo. \\ \\ No assumption on neighbour particles'  \\ density or velocity distribution} & \quantities{Relaxation of the hypothesis \\ $ \rm m \ll M_{BH}$ \\ \\ No assumption on neighbour particles'  \\ density or velocity distribution} & \quantities{
    $\rm b_{max} = \epsilon_{BH}$ instead of 10 ckpc/h 
   \\
   \\DF correction is not coupled \\ with a boost  \\ of the BH particles' dynamical mass.} \\
        \midrule
        \midrule
                \hline
        \end{tabular}
 \tablebib{(1) \citet{hirschmann_cosmological_2014}; (2) \citet{tremmel_off_2015}, (3) \citet{Chen}.}
\end{table*}

\subsubsection{Repositioning BH particles} \label{pinning}
Among the major issues encountered when introducing BH particles in N-body simulations is that they can escape from the centres of the host galaxies. One of the most widely used methods to avoid this consists of repositioning, at each time step, the BH particle at the position of the most bound particle among its neighbours. Different implementations of this method feature different choices for the search radius. Moreover, such alternative implementations often adopt additional constrains (e.g. on their relative velocity) for the selection of the neighbour particle on which to relocate the BH.
In this way, the BH particle is generally forced to remain at the centre of its host sub-halo \citep[see e.g.][]{Di_Matteo_2008, booth2009cosmological, vogelsberger2013model, Sijacki_2015, Schaye_2014, Pillepich_2017, Ragone_Figueroa_2018, Bahe2022}. 

As pointed out by \cite{tremmel_off_2015}, this method may have major shortcomings during mergers or high-speed close encounters between galaxies. During a close encounter between two galaxies, one of the two BHs may select the most bound neighbour particle as a particle belonging to the other galaxy. In this case, at the next time step the BH is suddenly and unphysically relocated to the neighbouring galaxy, thus leaving its original host galaxy without a central BH. In addition, this BH, which has typically moved to an outer region of the galaxy, will be quickly repositioned closer to the centre of the new host galaxy, where another BH is located, within a few time steps. In this way, BH-BH mergers will become faster and more frequent.

To prevent the occurrence of these spurious behaviours, different definitions of the neighbours, over which to search for the most bound particle, have been employed. The original radius was set as the SPH smoothing radius of the BH particle, or some kernel radius associated with a different hydro solver \cite[e.g.][]{Di_Matteo_2008, vogelsberger2013model, dave_simba_2019}. Other authors preferred to search the most bound particle within the BH gravitational softening or a small multiple of it \cite[e.g.][]{booth2009cosmological, Schaye_2014, Ragone_Figueroa_2018} because the smoothing length can be much larger, thus exacerbating the problem mentioned above.

A further condition usually introduced to search for the most bound particle is that its velocity relative to the BH has to be smaller than a threshold in the attempt to ensure that it belongs to the same galaxy. The most commonly used threshold is a fraction of the local sound speed, as originally introduced by \cite{Di_Matteo_2008}. However, \cite{Ragone_Figueroa_2018} found that this criterion is not effective, besides having a not-so-clear physical basis \cite[see also][]{Bahe2022}. Their results improved by imposing a maximum velocity of the order of the typical motions of particles within galaxies, namely 100-200 km/s. These values and the smaller search radius limited the unphysical transfer of a BH to another galaxy in their zoom-in massive cluster simulations since the typical orbital velocities of galaxies in clusters are much larger.
While including such additional criteria improves the performance of the method, in the simulation presented in this work we adopt a version of the repositioning closer to the original one, where the most bound neighbour particle is selected within the SPH smoothing length, provided that its relative velocity is smaller than 25 \% of the local sound speed. The motivations behind this choice lies in the increased resolution of the simulations presented here, which provide more complex merger dynamics, and the purpose to highlight the limitations of this commonly used repositioning technique.
However, we keep the BH velocity calculation unchanged, regardless of the specific sub-resolution approach used for BH dynamics.

\subsubsection{Boosting the dynamical mass} \label{dynamicalmass}
An alternative approach to account for the limited mass resolution when the mass of the BH particles is close to its seeding mass is to increase the BH dynamical mass at seeding artificially. In this approach, once seeded, two different masses are assigned to the BH: the real mass, $m_r$, which grows continuously by the Eddington-limited Bondi-like prescription (see Sect. \ref{accretion} below), and the dynamical mass, $m_d$, which enters in the computation of gravitational force. At seeding, the latter is set at a relatively large value, typically equal to the mass of DM particles, while the former is a few orders of magnitude smaller. As long as $m_r<m_d$, the value of $m_d$ does not increase by gas accretion, which only affects the value of $m_r$. Once $m_r\ge m_d$, the real mass increases in a continuous way, while the dynamical mass increases according to a stochastic prescription for the swallowing of neighbour gas particles \citep[e.g.][]{springeldimatteo}. From then on, the two masses remain similar, differing only because of the stochastic swallowing of gas particles. 

This artificial boost of the BH dynamical mass at seeding is intended to amplify the effect of the numerically resolved DF, eventually preventing the BH from escaping the host halo soon after it is seeded. 

Clearly, this method is  also prone to spurious effects. For instance, \cite{tremmel_off_2015} pointed out that initialising the BH mass with a value hundreds of times higher than its real value can affect the mass of the host galaxy, thus unavoidably impacting its subsequent evolution.

\vspace{5 mm}
In the present section, we review the most commonly used techniques to deal with BH dynamics, mainly to provide a background to the simulations presented in this work. We refer the reader to \cite{dimatteo2023massive} for a more comprehensive discussion.

\section{Super-massive black holes in cosmological simulations} \label{GeneralBHs}

In this section, we discuss the description of SMBH evolution and of the ensuing AGN feedback in our simulations. In Sect. \ref{og}, we briefly give an overview of the \OG code, within which we implemented our model to correct for unresolved DF. Given the finite force and mass resolution of cosmological simulations, sub-resolution models are needed to describe the processes of birth, accretion, and feedback of SMBHs. Furthermore, the merger events between BH pairs cannot be followed down to the final inspiraling of their orbits. Therefore, we need to include also some criteria to establish when to merge two BHs. In Sect. \ref{bh_cosmo} we review the approach to treat BHs in cosmological simulations, by focusing on the seeding criterion (Sect. \ref{seeding_sec}), on gas accretion (Sect. \ref{accretion}), and on the conditions allowing BH-BH mergers (Sect. \ref{mergers}).

\subsection{The \OG simulation code}
\label{og}
Our simulations are based on the \OG\ code (Dolag et al.,, in prep.; see also \citealt{groth.etal.2023}), which represents an evolution of the {\tt GADGET-3} code (which, in turn, is an improvement of the previous {\tt GADGET-2} code by \citealt{Gadget2_2005}). \OG\ solves gravity with the Tree-PM method \citep[see also][]{2016pcre.conf..411R}. In the simulations presented here, hydrodynamics is described by the SPH formulation presented by \cite{beck.etal.2016}, which overcomes several of the limitations of the original SPH formulation of GADGET-3. 

\OG\ is parallelised using a hybrid MPI/OpenMP/OpenACC scheme \citep{ragagnin.etal.2020}. Adopting a limited number of MPI tasks per node allows us to reduce the
‘communication surface', while efficiently using OpenMP inside a single shared-memory node. Load-balancing is achieved using a domain
decomposition based on a space-filling Peano-Hilbert curve, whose fragmentation into segments (each assigned to an MPI task) guarantees a
very good computational balance, at the expense of some memory imbalance \citep{Springel_2005}.

\OG\ includes a sub-resolution description of a range of astrophysical processes relevant for the simulations presented here: metallicity-dependent radiative cooling \citep[e.g.][]{wiersma.etal.2009}, an effective model for star formation from a sub-resolution description of the multi-phase structure of the interstellar medium \citep{springel.hernquist.2003}, a model for stellar evolution and chemical enrichment from AGB stars and type-Ia and II supernovae \citep{tornatore.etal.2007, Bassini_2020}, and a model to follow the evolution of SMBHs and the ensuing AGN feedback (see below for details). As for the latter, we remind that the aim of this paper is to present an improved implementation of a sub-resolution description of the effect of DF on the dynamics of BH particles (see Sect. \ref{Dynamics} for details).
\subsection{Black holes in cosmological simulations}
\label{bh_cosmo}
\subsubsection{Black hole seeding} \label{seeding_sec}
In our simulations, BHs are described by collisionless sink particles which are initially seeded within a halo hosting a ‘bona fide' galaxy. The halo is identified through a Friend-of-Friend (FoF) algorithm (with linking length equal to 0.2 times the mean separation of DM particles). For a BH particle to be seeded, we require the host halo to fulfill few conditions, so as to guarantee that it is well resolved and that star formation already took place within it. Following \cite{hirschmann_cosmological_2014}, we added to the halo mass threshold criterion introduced by \cite{Springel_2005} additional conditions for star and gas fraction of the halo.

In detail, in the simulations presented in this work, the following seeding conditions must all be met: \emph{(i)} the DM mass of the halo exceeds the value of $6.94 \times 10^{10}$ M$_\odot$; \emph{(ii)}  the stellar mass is at least 2 per cent of the total mass and 5 per cent the DM mass of the halo; \emph{(iii)} the gas mass reaches a value of 10 per cent of the stellar mass; \emph{(vi)} the halo does not contain any other BH particle. If a halo fulfils these conditions, the most bound star particle of the halo is converted into a BH. The mass of a BH particle at seeding, $M_{\rm BH,seed}$, is not fixed, but scales with the amount of stars in the FoF group according to:
\begin{equation} \label{seeding}
M_{\rm BH,seed} = M_0 \, \frac{M_{\rm *,h}}{f_* \,M_{\rm DM,seed}}\,,
\end{equation}
where $M_{ \rm *,h}$ is the stellar mass assigned to a FoF halo, $f_*$ and $M_{{\rm DM,seed}}$ are the input parameters for the fraction of stellar mass and the minimum DM mass for a FoF halo where to seed a BH, respectively. As for $M_0$, it is the minimum seeding mass of a BH particle, when the seeding conditions are just met.
{In the simulations presented in this work, the mass of star particles (see Sect.\ref{Simulations} and Table \ref{tab:BHdetails} ) is larger than $M_0$.
Therefore, the assumption that the BH mass is larger than the one of the surrounding objects}, on which the derivation of the Chandrasekhar DF formula is based, is not met at seeding due to limited mass resolution, and for the above condition to hold BHs need to grow by a large factor.
Therefore, it is not surprising that, despite the initial BH position is at the location of the most bound particle, two-body scatterings with neighboring particles can easily cause the BH to be scattered outside its host galaxy.
\subsubsection{Gas accretion onto BHs} \label{accretion}
During its evolution, the mass of a BH increases through two channels: accretion of surrounding gas and merging with other BHs. As for the former, we adopt the accretion model originally implemented by \cite{Springel_2005}. 
BH accretion rate is calculated according to Bondi-Hoyle formula \citep{hoyle, bondi44, bondi52} as

\begin{equation} \label{bondi}
{\dot M_{\rm Bondi}} = \frac{4 \pi G^2 M^2_{\rm BH} \alpha \rho}{(c_s^2 + v^2)^{3/2}}\,,
\end{equation}
where $\rho$ and  $c_s$ are the density and the sound speed of the surrounding gas computed at the position of the BH particle, $v$ is the relative velocity of the BH with respect to the surrounding gas particles, and $\alpha$ is a ‘boost' factor introduced to account for the limited resolution with which gas density in the surroundings of the BHs is reconstructed. Following \cite{steinborn.etal.2015}, we distinguish between accretion from the hot ($\mathrm{T>10^{5}}$ K) and the cold gas  ($\mathrm{T<10^{5}}$ K), using $\alpha=10$ and $\alpha=100$ respectively for the hot and the cold gas.

The accretion is always limited to the Eddington accretion rate,

\begin{equation} \label{eddingtonrate}
\dot{M}_{\rm Edd} = \frac{4 \pi G M_{\rm BH} m_p}{\eta_r \sigma_T c}\,,
\end{equation}
where $m_p$ is the proton mass, $\sigma_T$ is the Thompson cross-section. The parameter $\eta_r$ is the radiative efficiency and represents the fraction of the accreted rest-mass energy which is converted in radiation  \citep{novikov1973astrophysics, noble2011radiative}. In our simulations, we use $\eta_r=0.1$.
In addition, we allowed the black hole to swallow gas particles via stochastic accretion, as originally proposed by \cite{Springel_2005}. Following \cite{fabjan.etal.2010}, gas particles are not swallowed entirely, but are sliced into three parts, so as to have a more continuous description of stochastic accretion. 
In this way, we can assign to each BH two masses that in general slightly differ: a ‘true' mass, which grows in a continuous way according to the above Eddington-limited Bondi-like accretion model, and a dynamical mass, which is varied each time that a portion of a gas particle is stochastically selected for the swallowing. 
\subsubsection{Mergers} \label{mergers}
The finite  force resolution set by the gravitational softening sets the scale below which gravitational interactions, including mergers between BHs, cannot be properly followed. 
The simplest criterion for defining when a BH-BH merging event occurs is to impose a limiting BH-BH distance, $d_{\rm merg}$, below which the two BHs could
immediately merge. In our simulations, we adopt a value of $d_{\rm merg}= 5 \cdot \epsilon_{\rm BH}$.
However, during fly-by encounters between galaxies, it could happen that this criterion produces an unwanted behaviour, with the two BHs forced to merge, even if their relative velocities are large enough to make them gravitationally unbound. 

To circumvent this potential problem, we  include in our simulations two additional criteria that must be fulfilled for a merger to happen. Following similar arguments as in \cite{Di_Matteo_2008}, we first require that 
\begin{equation}
    v_{ \rm rel} < 0.5 \ c_s\,.
\end{equation}
Furthermore, we impose that the two BHs are gravitationally bound by requiring:
\begin{equation} \label{mergercond_phi}
\frac{|\Phi_{\rm BH_1}-\Phi_{\rm BH_2}|}{a} < 0.5 \cdot c_s^2 - v_{\rm rel}^2\,,
\end{equation}
where $\Phi_{\rm BH_1}$ and $\Phi_{\rm BH_2}$ are the values of the gravitational potential at the positions of the two merging black holes, $c_s$ is the local sound speed, $v_{\rm rel}$ their relative velocity and $a$ is the scale factor, normalised to unity at redshift $z=0$.
During a merger event involving two BHs, the BH having a lower value of the potential swallows the other BH. Consequently, after the merger, the surviving BH retains its original position and velocity, while its mass increases by the mass of the swallowed BH.

\subsubsection{Feedback energy} \label{feedback}
As a BH accretes gas, it injects energy in the surrounding region and a fraction of the energy radiated during gas accretion is thermally coupled in the form of feedback energy to the surrounding medium \citep[for an extensive comparison of different implementations of AGN feedback in simulations]{Wurster_2013}.
Within the simulations presented in this work, we consider a purely thermal mechanism of feedback, whose energy rate is calculated according to
\begin{equation}
\dot{E}= \eta_r \eta_f \dot{M}_{\rm BH}c^2\,.
\end{equation}
In the above expression, $\dot M_{\rm BH}=\min(\dot M_{\rm Bondi},\dot M_{\rm Edd})$ is the BH mass accretion rate, and $\eta_f$ is the fraction of the radiated energy that is thermally coupled to the surrounding gas. 
Furthermore, we emulate a transition between the radio and quasar mode of BH feedback by varying the parameter $\eta_f$, as described by \cite{Sijacki_2006} and \cite{fabjan.etal.2010}. Whenever the accretion rate of the BH is one-hundredth of the Eddington limit, we increased the fraction of energy thermally coupled with the surrounding gas by a factor of four. We adopt a value of $\eta_f = 0.05$ during the quasar mode, increasing to $\eta_f = 0.2$ during the radio mode. 
\hspace{20 mm}

For the sake of clarity, we summarise in Table \ref{tab:BHdetails} the information on the force resolution and main characteristics of BH model implemented in the simulations performed for this work, which are described in the next section. 
\begin{table*}
        \centering

        \caption{Relevant equations and parameters that characterise the sub-resolution model of BH evolution and AGN feedback. }
    \label{tab:BHdetails}
        \begin{tabular}{p{3.2cm} p{3.5cm}p{3.5cm}p{3.3cm}p{3cm} }
                \hline
        \toprule 
                Resolution & Seeding & Accretion & Merger&  Feedback \\
        \hline
                \midrule
                  \quantities{ {Mass} \\ 
          $m_{\rm DM} = 4.69 \times 10^{7}$ M$_\odot$ \\ 
          $ m_{\rm gas} \  = 8.67 \times 10^{6}$ M$_\odot$ \\
          $ \ \  m_{{\rm *}}   \ \  = 8.7 \times 10^{6}$  M$_\odot$ \\ ~\\
          {Force softening} \\
          $\epsilon_{ \rm DM} = 4.13 $ ckpc / $1.38$ kpc 
          \\
          $\epsilon_{\rm gas} = 1.38 $ ckpc
          \\
          $\epsilon_{\rm *}  \  \ = 0.35 $ ckpc
          \\
           $\epsilon_{\rm BH} \ = 0.35$ ckpc }&
          \quantities{ $ M_{{\rm BH,seed}}=M_{\rm 0} \frac{M_{\rm *,h}}{f_* M_{\rm DM,seed}} $ \\ \\
          
          $M_{\rm 0} = 2.7 \times 10^{5} $ M$_\odot$ \\
           \\ $M_{{\rm DM,seed}}=6.94 \times 10^{10}$ M$_\odot$ \\  \\
          $\frac{M_{\rm *,h}}{M_{\rm {h}}}=2\%$ \\  \\$\frac{M_{\rm *,h}}{M_{\rm DM,h}}=5\% $ \\   \\ $\frac{M_{\rm gas,h}}{M_{{*,h}}}=10\% $}
            &  
            \quantities{$\min( \dot{M}_{\rm Bondi}, \dot{M}_{\rm Edd})$ \\
            ${\dot M_{\rm Bondi}} = \frac{4 \pi G^2 M^2_{\rm BH} \alpha \rho}{(c_s^2 + v^2)^{3/2}}$   \\   
            $\dot{M}_{\rm Edd} = \frac{4 \pi G M_{\rm BH} m_p}{\eta_r \sigma_T c}$ \\
            $\alpha = 10/100$ \\
            $\eta_r = 0.1$ \\ } 
            
            &
\quantities{Threshold distance: \\ $\Delta r_{\rm BH} < d_{\rm merg}$  \\  $d_{\rm merg} = 5 \cdot \epsilon_{\rm BH}$ \\

            \\ Gravitationally bound:
            \\ $v_{\rm rel} <  0.5 \cdot c_s $
            \\$\frac{|\Phi_{\rm BH_1}-\Phi_{\rm BH_2}|}{a} < 0.5 \cdot c_{\rm s}^2 - v_{\rm rel}^2 $ }  & \quantities{ $\dot{E}_{\rm feed}= \eta_{\rm r} \eta_{\rm f} \dot{M}_{\rm BH}c^2$
            \\
            \\
            $\eta_{\rm r} = 0.1$
            \\ \\ Quasar : $\eta_{\rm f} = 0.05$ \\  \\ Radio: $\eta_{\rm f} = 0.2$} 
            \\
                
        \midrule
                \hline
  \label{tab: simulations}
        \end{tabular}
 \tablefoot{Column 1: values of the masses of the different particle species in the high-resolution regions of \texttt{Dianoga} clusters and \texttt{Cosmobox}: $m_{ \rm DM}$, $m_{\rm gas}$ and $m_{*,\rm}$ are the masses of DM, gas and star particles, respectively; $\epsilon_{\rm DM}, \epsilon_{\rm gas}, \epsilon_{*}$, and $\epsilon_{\rm BH}$ are the Plummer-equivalent  softening lengths of the different particle species. For BH, gas and star particles, the softenings are fixed in comoving units. For DM particles, we fix the softening in comoving units until $z=2$, and then in physical units until $z=0$.  Column 2: criteria for BH seeding: $M_{\rm BH, seed}$ is the seeding mass of a BH; $M_{\rm 0}$ is the minimum BH mass seed; $M_{\rm DM,h}$, $M_{\rm gas,h}$ and $M_{\rm *,h}$ are the halo masses in the DM, gaseous and stellar components; $M_{ \rm h}$ is the total halo mass and $M_{\rm h,seed}$ is the minimum halo mass required to seed a BH. Column 3: BH mass accretion rate: $\dot{M}_{\rm Bondi}$ and $\dot{M}_{\rm Edd}$ are defined in Eqs.\ref{bondi}, \ref{eddingtonrate},. Column 4: criteria for BH-BH merging: $\Delta r_{\rm BH}$ is the distance between the merging BHs, $d_{\rm merg}$  is the minimum threshold distance required for merging, $\Phi_{ \rm BH_1}$ and $\Phi_{\rm BH_2}$ are the values of the gravitational potentials of the two BHs, $v_{\rm rel}$ the magnitude of their relative velocity, $c_{\rm s}$ the local sound speed and $a$ is the scale factor. Column 5: expression of the rate of the feedback energy released by the BH and thermally coupled to the surrounding gas; $c$ is the light speed, $\eta_{\rm r}$ and $\eta_{\rm f}$ are the radiative and coupling efficiency, respectively. See Sect.\ref{GeneralBHs} for additional details. }
\end{table*}
\section{Simulations} \label{Simulations}

\begin{table}
        \centering
        \caption{Set of the simulations.}
        \label{tab: simulations}
        \begin{tabular}{lccr} 
                \hline
        \toprule 
                Simulation & Name &  $M_{\rm vir}$ [$10^{13}$ M$_\odot$] &  $\rm R_{\rm vir} $ [kpc] \\
        \hline
                \midrule
                 & \texttt{REPOS}   & 1.27 & 610.89 \\
                 \texttt{Cl13}& \texttt{DYNMASS} & 1.26 & 609.33\\
         &  \texttt{DYNFRIC}  & 1.26 & 609.75 \\
        \midrule
        & \texttt{REPOS} &  18.76& 1497.17   \\
         \texttt{D9}& \texttt{DYNMASS}& 19.02 &  1504.28 \\
          & \texttt{DYNFRIC} &  18.89  & 1499.24  \\
        \midrule
         & \texttt{REPOS} & 2.57 &  751.01 \\
         \texttt{CosmoBox}& \texttt{DYNMASS} & 2.58 & 751.55  \\
         
          & \texttt{DYNFRIC} & 2.65   &   760.15 \\
        \\
        \midrule
                \hline
        \end{tabular}
 \tablefoot{ Column 1: simulation name; Column 2: method for accounting for unresolved DF; Column 3 and 4: virial mass and virial radius, respectively, at $z=0$, of the most massive halo in the  simulations}
\end{table}
To assess the performance of the novel DF correction and to compare it with other prescriptions (Sect.~\ref{others}), we carried out a set of two zoom-in simulations (a group-sized and a cluster-sized halo), and a cosmological box (with side length of 16 comoving-Mpc), all at the same resolution. This allowed us to test the new implementation in different environments. 
All simulations are performed assuming a flat $\Lambda$CDM cosmology with $\Omega_m = 0.24$, $\Omega_b = 0.0375$ for total matter and baryon density parameters, $h=0.72$ for the Hubble parameter, $n_s =0.96$ for the primordial spectral index, $\sigma_8 =0.8$ for the power spectrum normalisation.
We report in Table \ref{tab: simulations} a description of the main characteristics of these simulations. 
The selected zoom-in regions belong to the {\tt{Dianoga}} simulation set introduced by \cite{Bonafede_2011} and \cite{Bassini_2020}. We refer to such papers for a detailed description of this set of zoom-in initial conditions for simulations of galaxy clusters, as well as for details on the model of star formation and chemical enrichment adopted. The \texttt{D9} cluster has a mass $M_{\rm 200}\simeq 1.53\times 10^{14}$ M$_\odot$ at $z=0$, while the other refers to a lower-mass group of M$_{\rm 200}\simeq 1.38 \times 10^{13}$ M$_\odot$, namely \texttt{Cl13}\footnote{We indicate with $M_{\rm 200}$ the total mass contained within a sphere of radius $\rm R_{\rm 200}$ which contains an overdensity equal to $200 \rho_c(z)$, where $\rho_c(z)= 3 H(z)^2/8\pi G$ is the critical cosmic density at redshift $z$. In Table \ref{tab: simulations} we use the ‘virial radius'  $\rm R_{\rm vir}$ defined as the radius within which the mean halo density corresponds to the prediction of spherical collapse \citep[e.g.][]{eke.etal.1996}. Accordingly, the ‘virial mass' $\rm M_{\rm vir}$ is the mass of a sphere enclosed within a radius $ \rm R_{\rm vir}$.}. 
Both objects have been simulated at the same mass and spatial resolution (see Table \ref{tab:BHdetails}).

As for the cosmological box, namely \texttt{CosmoBox}, it has been simulated by adopting exactly the same cosmological model and the same implementation of all physical processes as the zoom-in simulations. 
For each initial condition of \texttt{D9}, \texttt{Cl13}, and \texttt{CosmoBox}, we carried out three simulations using the same settings while changing only the sub-resolution technique to cope with the BH dynamics. We compared the repositioning scheme, the adoption of a boosted dynamical mass and the new implementation of DF introduced in  Sect.\ref{Dynamics}. 
In the following sections, we refer to these schemes as, respectively, \texttt{REPOS}, \texttt{DYNMASS} and \texttt{DYNFRIC}.
Comparing the results of these simulations allows us to assess the effect of our new implementation of DF with respect to some previously adopted ad hoc prescriptions to account for this effect.

In the following sections, we present our results using two different approaches. In Sect. \ref{GlobalProperties}, we present a statistical analysis to show how the method used to track BH orbits impacts the properties of their population. Section \ref{zoom} is dedicated to the study of the evolutionary dynamic histories of individual BHs, comparing their path when they sink into the potential well of the host sub-halos and during mergers when governed by different sub-resolution prescriptions.
\section{Properties of the BH population} 
\label{GlobalProperties}

The minimum aim of our DF model is to avoid spurious numerical effects that other methods may produce, as discussed in Sect. \ref{Dynamics}. Thus, in this section, we present an analysis of the overall properties of the population of SMBHs in our simulation, focusing on the ability of our DF model to: place and hold the BH at the centre of the host galaxies (Sect. \ref{centring}), prevent BH particles from spuriously wandering outside the host galaxies (Sect. \ref{wandering}), reproduce the co-evolution of BHs and host galaxies (Sect. \ref{magorrian}).

\subsection{Centring the BHs within host galaxies} \label{centring}
To assess the ability of each model to correctly locate the BHs at the centre of the host sub-halos, we select all the main halos and sub-halos having a BH within their DM half-mass radius (hereafter $\rm R_{\rm HMS}$). 
We associated each BH to the closest (sub-)halo, as this procedure allows us to assess how BHs are kept at the centre of their host or are off-centred.
Sub-halos are identified by the {\tt{SubFind}} algorithm \citep{Springel.etal.2001, subfind}  and are requested to feature at least 20 particles. In the zoom-in simulations, we excluded from the analysis all the halos that contain within their $\rm R_{200}$ at least one low-resolution DM particle spuriously scattered within the high-resolution region. We then calculated the distance between the (sub-)halo centres and the closest BH. 
The sub-halo centres are identified with the minimum of the gravitational potential occurring among the member particles of that sub-halo. Their values are retrieved from the {\tt{SubFind}} output. 

\begin{figure}

\includegraphics[scale=0.60]{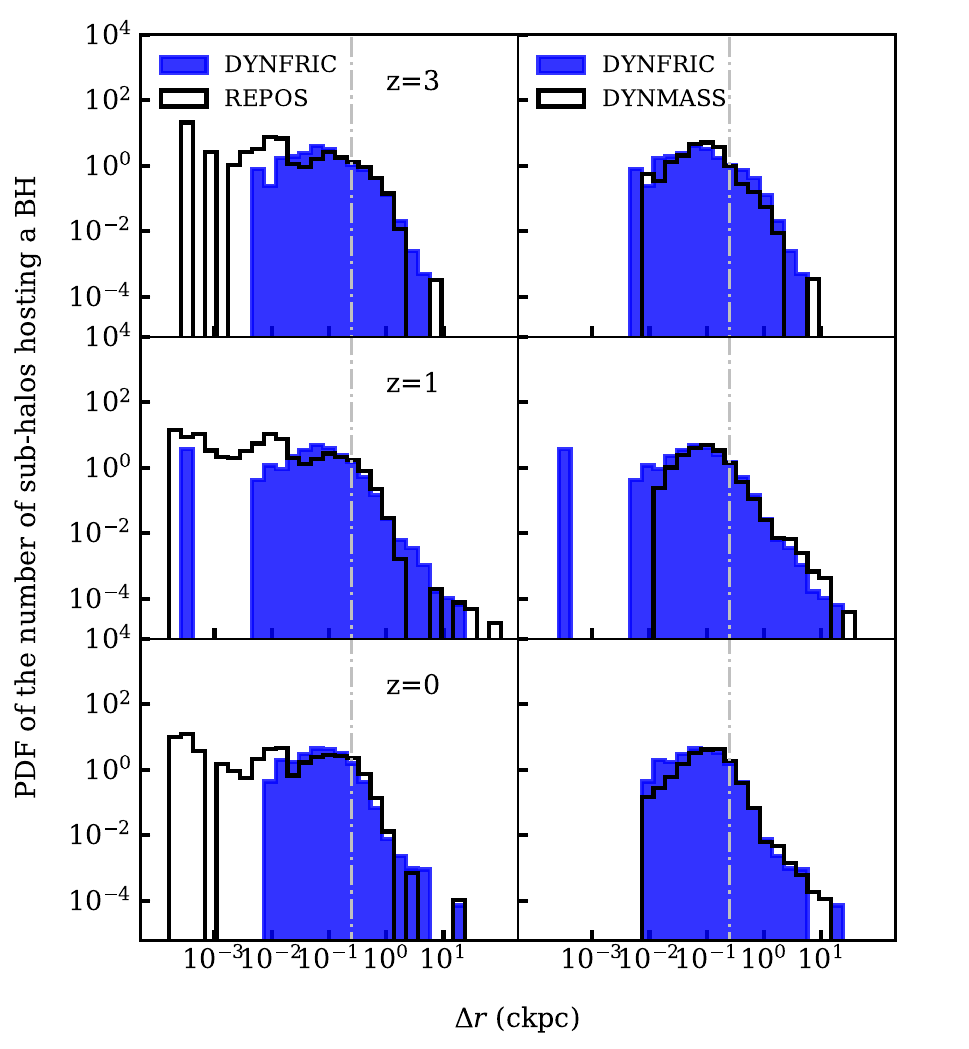}

\caption{Probability density distribution  of the distances between sub-halos identified by {\tt{SubFind}} and the closest BH particle within the  $\rm R_{\rm HMS}$ of each sub-halo of \texttt{Cl13}, \texttt{D9} regions, and \texttt{Cosmobox}.
 The rows show the results obtained at different redshifts: $z=3$ (up), $z=1$ (central), and $z=0$ (bottom). The column report the results comparing \texttt{DYNFRIC} and \texttt{REPOS} on the left, and \texttt{DYNFRIC} and \texttt{DYNMASS} on the right side. We include a dashed-dot line in each plot indicating the softening length of the BH as a reference for the spatial resolution of the simulation.}
\label{fig:distancestot}
\end{figure}

Figure \ref{fig:distancestot} shows the  histogram of the probability density function (PDF) of the number of (sub-)halos having a BH within their  $\rm R_{\rm HMS}$, versus the distance $\Delta r$ from the associated BH. The sample of sub-halos in Fig. \ref{fig:distancestot} is obtained by summing up all the (sub-)halos from \texttt{D9}, \texttt{Cl13} and \texttt{CosmoBox}. However, we verified that the individual analysis of each of these simulations keep the overall results unchanged.
In each column, we compare different prescriptions for BH dynamics with the DF model, shown in blue, and we report the results at $z=3$, 1 and 0, from top to bottom.  For reference, the dashed-dot vertical line marks the value of the Plummer-equivalent softening length of BH particles.
Not surprisingly, the \texttt{REPOS} prescription predicts sub-halos that host generally well-centred BHs. 
This is the most consistent advantage of the repositioning technique, as well as the most predictable, since it explicitly forces the BHs to be relocated at each time-step at the centre of their host substructures.
As for the \texttt{DYNMASS}, it shows the less pronounced tail for higher distances ($\Delta r>1$ ckpc) at $z=3$. In fact, at high redshift, the impact of a large dynamical mass is relatively more important given that a larger number of BHs is expected to have relatively smaller true mass and live in less resolved hosting halos. Interestingly, in this regime, our DF model performs quite well in keeping BHs at the centre of their host halos, without resorting to ad hoc prescription. The comparable performance of the two methods is actually a nontrivial and encouraging result for our DF implementation; in fact, accurate centring of the BHs in sub-halos is obtained as a result of an implementation of the DF correction instead of an artificial increase of BH masses, which have the side effect of altering the structure of the sub-halos gravitational potential, thereby impacting on the resulting galaxy formation process \citep[e.g.][]{Chen}.
Table \ref{tab:percent_cent} reports the percentage of sub-halos with which each simulation contributes to the differential distribution of Fig.\ref{fig:distancestot}. The \texttt{D9} simulation provides the most numerous population of centred sub-halos, with at least $70 \%$ of the total distribution.

\begin{table*}[]
\caption{Percentages of sub-halos from each simulation that contribute to the distributions shown in Fig.\ref{fig:distancestot}. }
\label{tab:percent_cent}
    \centering
\begin{tabular}{*{10}{c}}
    \hline
    \toprule
    & \multicolumn{3}{c|}{REPOS}
    & \multicolumn{3}{c|}{DYNMASS}
    & \multicolumn{3}{c}{DYNFRIC} 
            \\
  &   \texttt{Cl13}  &   \texttt{D9}  &   \texttt{CosmoBox} &   \texttt{Cl13}  &   \texttt{D9}  &   \texttt{CosmoBox} &   \texttt{Cl13}  &   \texttt{D9}  &   \texttt{CosmoBox}   \\
    \hline
    \hline
$z=3$   &   11  &   82  &   7  &   12  &   83  &   5  &   12  &   83 & 5  \\

$z=1$   &   10  &   78  &   12  &   9  &   80  &   11  &   10  &   78 & 12  \\

$z=0$   &  13  & 70 & 17  & 10 &75   &  15     &   10    &  75 & 15     \\
\hline
\hline
\end{tabular}
\tablefoot{For each implementation, we indicate the percentage of sub-halos for each simulation: \texttt{Cl13}, \texttt{D9} and \texttt{Cosmobox}, in the left, central and right columns, respectively. Percentages are listed for different redshifts (as in Fig.\ref{fig:distancestot}): $z=3$, $z=1$, and $z=0$.}
\end{table*}

\subsection{The population of wandering BHs} \label{wandering}
Another crucial aspect to consider when judging the reliability of a DF model is its capability of preventing BHs from escaping the host sub-halos because of two-body interactions with larger-mass particles or during merger events. We denote the population of these kicked-off BHs as wandering BHs.
We classify a BH as wandering whenever its distance from the closest sub-halo is larger than twice the  $\rm R_{\rm HMS}$ of the closest sub-halo.
With this definition of WBHs we aim to focus on those BHs that have been expelled from their parent host galaxies as a consequence of spurious numerical effects. Thus, they should not be confused with the ‘classical' population of off-centre WBHs as, for instance, investigated by \cite{Tremmel_2018_wandering}, \cite{Ricarte_2021}, which instead are expected to become wandering as a consequence of numerically resolved gravitational dynamics.
It is worth noting that our definition can misidentify wandering BHs during close encounters between sub-halos, when a non-WBH beloging to a sub-halo can temporary become a wander of the closer sub-halo passing by. We verified that limiting the search only to those BHs with a distance exceeding twice the $\rm R_{\rm HMS}$ of the closer among the two sub-halos, the results reported below do not vary considerably.

\begin{figure}
    \centering
    \includegraphics[scale = 0.33]{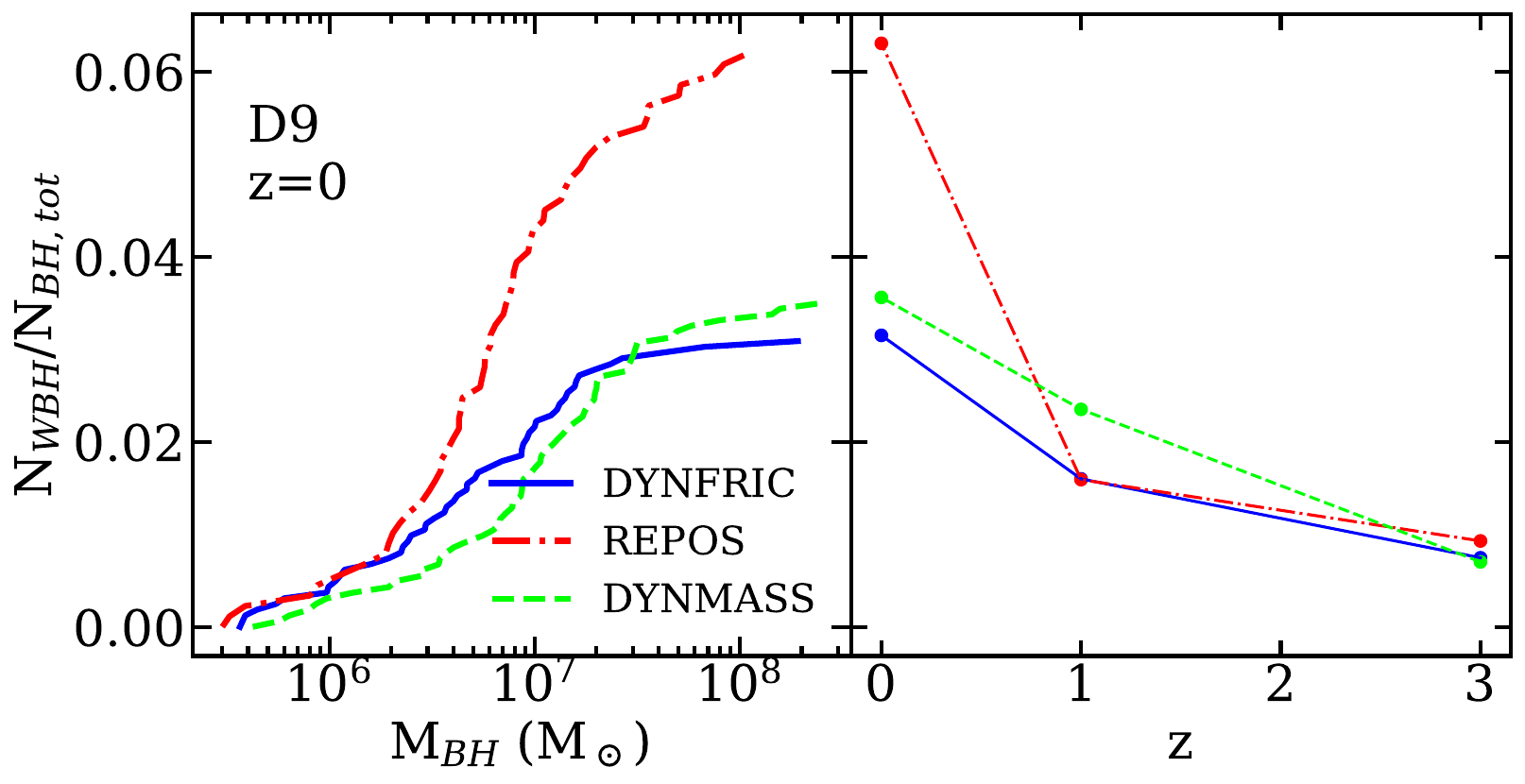}
    \caption{Cumulative number of wandering BHs to the total number of BHs found in  the \texttt{D9} region as a function of the BH mass at $z=0$ (left panel), and as a function of redshift for $z=3$, $z=1$, and $z=0$ (right panel). In the left panel, no mass threshold is adopted to select the BHs, and hence all the BHs with mass above the seeding mass are included (see Table \ref{tab:BHdetails}). In both panels, the simulation using the  DF model, marked with a blue line, is compared with \texttt{REPOS} in red and \texttt{DYNMASS} in green. The wandering BHs are defined as those having a distance larger than two times the $\rm R_{\rm HMS}$ from the closest sub-halo.}
    \label{fig:wbh}
\end{figure}
Throughout this section, we only present the results from the \texttt{D9} simulation. Its larger population of WBHs enables a more accurate statistical analysis.
\\

Figure \ref{fig:wbh} shows the cumulative fraction of wandering BHs to the total number of BHs as a function of BH mass at $z=0$ in the left panel and as a function of redshift in the right panel, for all the considered sub-resolution models. 
In the left panel, for each BH mass on the x-axis we see the cumulative fraction of wandering BHs on the y-axis. 
We observe that, irrespective of the method employed to keep the BH centred, the ratio of the wandering BHs to the total number of BHs increases as we progressively consider larger BH masses at the numerator. Each curve extends up to the value of the mass of the most massive wandering BH found in the considered simulation.
The \texttt{REPOS} simulation, in particular,  exhibits the higher fraction of wandering BHs. The percentage of wandering BHs in the simulation using the repositioning scheme rises for massive BHs ($\rm M_{\rm BH}>10^{7} \rm M_{\rm \odot}$).
\begin{figure}
    \centering
    \includegraphics[width=1\linewidth]{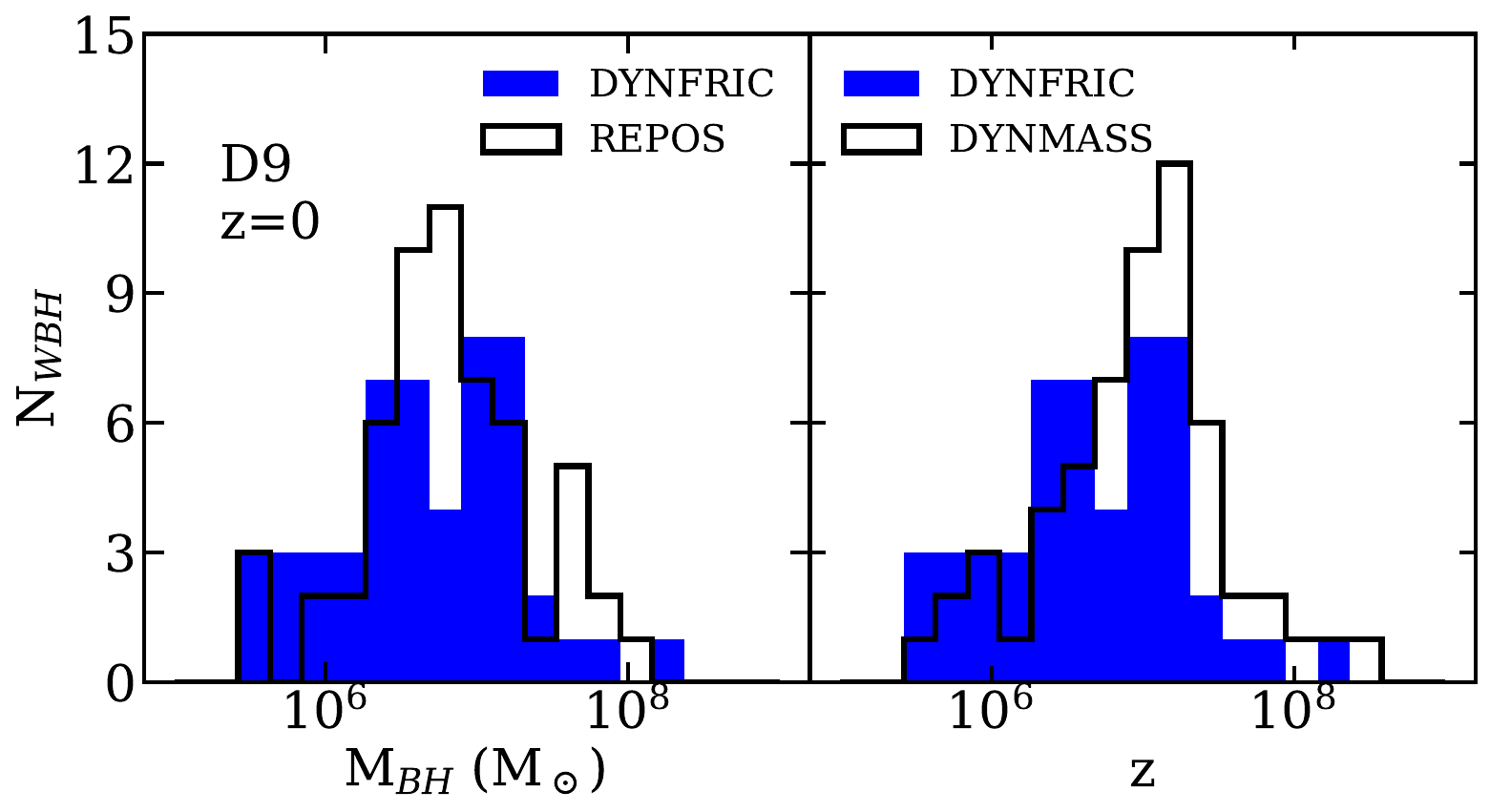}
    \caption{Histogram of the number of WBHs as a function of their mass for the \texttt{D9} simulation at z=0. We overplot the distribution of \texttt{DYNFRIC} and \texttt{REPOS} on the left and \texttt{DYNFRIC} and \texttt{DYNMASS} on the right.}
    \label{fig:noncumulativewbhs}
\end{figure}
Figure \ref{fig:noncumulativewbhs} shows the number of WBHs for each mass bin found in the \texttt{D9} simulation at $\rm z=0$. 
For every simulations, the majority of wandering WBHs' masses lies below $ \rm 10^{8}$. 
The \texttt{REPOS} simulations shows a peak at $\rm M_{BH} \sim 10^{7} M_\odot$. 
Interestingly, this implementation shows a second lower peak for $\rm 10^{7}<M_{BH}<10^8$. Again, this suggests that the repositioning technique adopted in this work can produce a significant displacement or BH ‘teleporting' of even massive BHs. The peak of the distribution in the \texttt{DYNMASS} simulation is shifted toward larger masses compared to \texttt{REPOS}. Moreover, \texttt{DYNMASS} reproduces lower WBHs for  $\rm M_{BH} < 10^{7}$ compared to \texttt{DYNFRIC} and more for $\rm M_{BH} > 10^{7}$. On one hand, BHs in the low-mass range are boosted in mass, preventing them from escaping the surrounding sub-halo. On the other, once their mass grows up to $ \rm M_{\rm DM}$, no correction acts on the dynamics of the BHs. Using the \texttt{DYNFRIC} technique, the peak of the distribution flattens compared to the other simulations. The number of WBHs with  $\rm M_{BH} < 10^{7}$ is  slightly higher than for $\texttt{REPOS}$ and $\texttt{DYNMASS}$, but it reduces in the high-mass end.

\begin{figure*}
\centering
    \includegraphics[scale = 0.65]{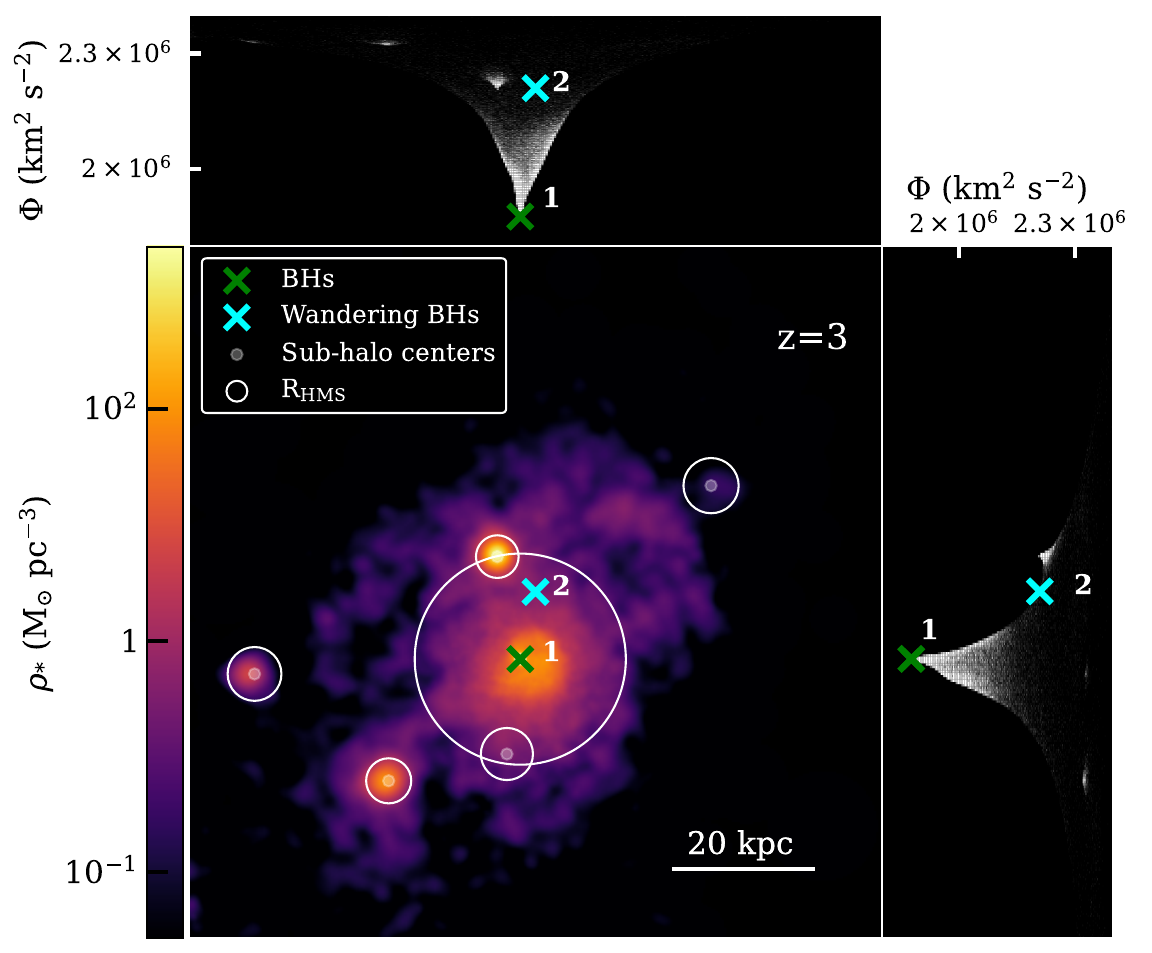}
        \includegraphics[scale = 0.65]{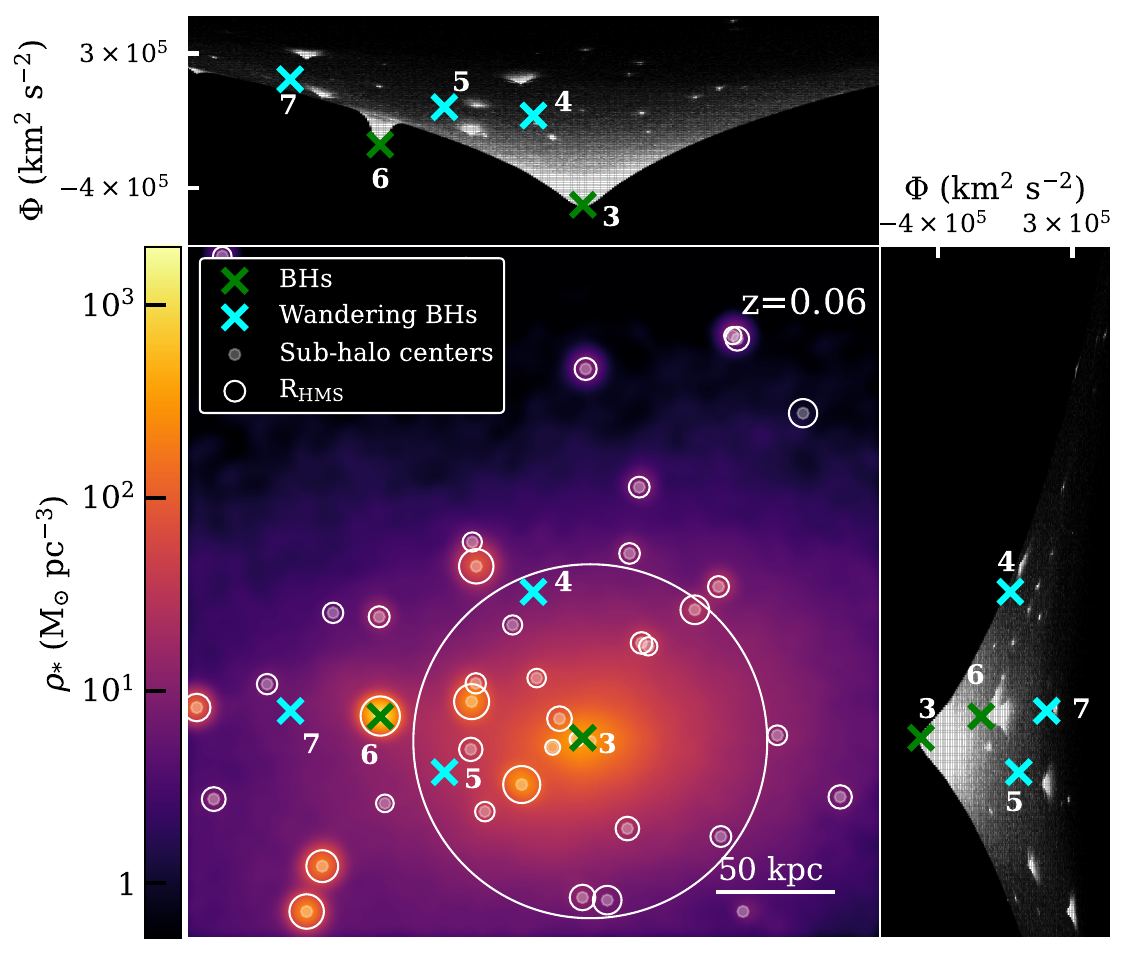}
        \caption{Stellar density projection along the $z$-axis with depths of $100$ kpc (top) and $300$ kpc (bottom), centred on the position of a wandering BH at redshift $z=3$ (tagged as BH \texttt{2}; left panel), and of a wandering BH at $z=0.06$ (tagged as BH \texttt{4}; right panel) in the \texttt{D9} simulation using the \texttt{REPOS} scheme. In both panels, the cyan crosses identify the wandering BHs, while the green crosses identify the BHs centred in their host. We show sub-halo centres as white shaded dots, with the white circles indicating the corresponding $\rm R_{\rm HMS}$. On the top and on the right of each panel we show the gravitational potential $\Phi$ of all the star particles (light white dots)  and of the BHs (crosses) in the field, projected along the two orthogonal directions. }
    \label{fig:wbh_events}
\end{figure*}

The increase of wandering BHs using the repositioning mechanism is surprising: these massive BH should reside in sub-halos having a deeper potential well. Therefore, they should be less prone to wander outside their host galaxy.
The origin for the excess of wandering BHs in the \texttt{REPOS} scheme can be ascribed both to close encounters between sub-structures and, at decreasing redshift, to the  presence of large-scale potential gradients. To illustrate how these mechanisms work, we show several wandering BHs in the two panels of Fig. \ref{fig:wbh_events}. 

Each panel of this figure shows, on the central plot, the projected stellar density around two wandering BHs, BH \texttt{2} in the top panel, and  BH \texttt{4} in the bottom panel. We also show the position of the other BHs in the field, marking wandering BHs with cyan crosses and the BHs centred on their host as green crosses. Furthermore, the centres of neighboring sub-halos are denoted by shaded white dots, with circles corresponding to the sub-halo $\rm R_{\rm HMS}$ enclosing them. The side plots display the gravitational potential $\Phi$ of the stars along the $x$-axis (on the top) and $y$-axis (on the right). 
In the top panel of Fig.\ref{fig:wbh_events}, BH \texttt{2} wanders due to a close encounter between two sub-halos, one having a deeper potential well. BH \texttt{2} then migrates to the location of the most bound neighbouring particles, thus moving along the gradient of the potential of the central sub-halo, and leaving its initial host sub-halo. It is worth noticing that, as defined here, a BH lying within twice the $\rm R_{\rm HMS}$ of a sub-halo can be a wander: our definition classifies as not wandering only those BHs located within twice the  $\rm R_{\rm HMS}$ of the closest sub-halo. 
The bottom panel of Fig. \ref{fig:wbh_events} shows the position at $z=0.06$ of the supermassive wandering BH \texttt{4} with a mass of $\rm 10^{10} \ M_{\odot}$. Being embedded in the large-scale potential gradient of the most massive halo, depicted in the side plots of the panel, BH \texttt{4} escaped from its original sub-halo.
This demonstrates that for the repositioning scheme implemented in this work, even the most massive BHs can move away from their host galaxies, thereby completely changing their gas accretion and the ensuing release of feedback energy. 

In the simulations using the DF correction and \texttt{DYNMASS}, wandering BHs instead originate from two-body or even three-body scattering. When the BH is not repositioned, the dynamics during these events can be quite complex; this aspect is analysed in more detail in Sect. \ref{zoom}.

\begin{figure*}
    \centering
    \includegraphics[scale = 0.8]{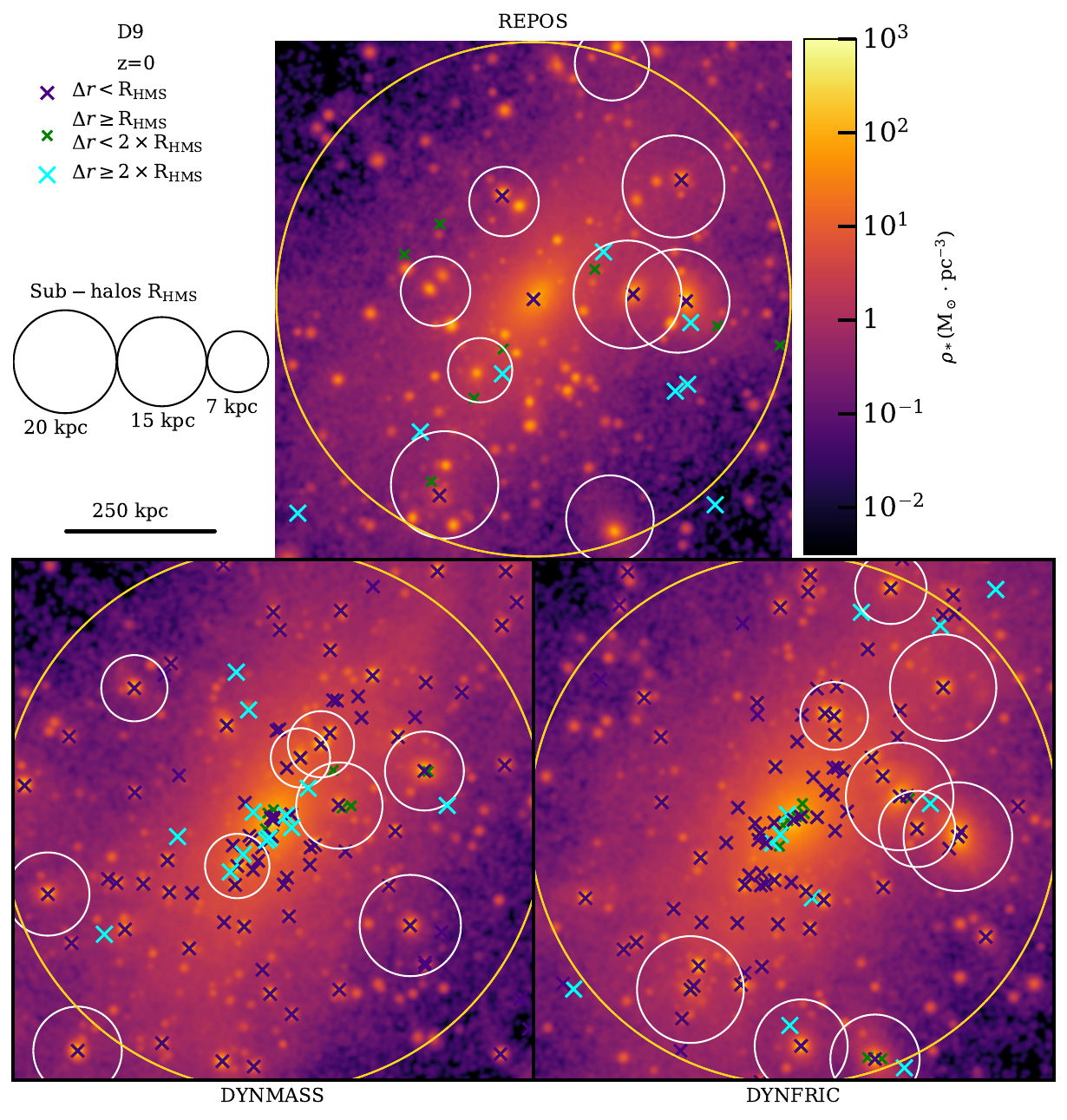}
    \caption{Stellar density maps along the $z$-axis with a depth of $1$ Mpc centred on the most massive halo of \texttt{D9} at $z=0$ in the \texttt{REPOS} (top), \texttt{DYNMASS} (bottom left), and \texttt{DYNFRIC} (bottom right) simulations. The panels are all 1 Mpc on a side.
    In each panel, we plot with dark-blue crosses the BHs located within the $ \rm R_{\rm HMS}$ of the associated sub-halo (the same criterion adopted in Sect. \ref{centring}). BHs lying between $ \rm R_{\rm HMS}$ and 2R$_{\rm HMS}$ of the closest sub-halo are indicated with green crosses, while wandering BHs (defined as those located beyond $2 \rm R_{\rm HMS}$ of the closest sub-halo) are shown as light-blue crosses. The values of R$_{\rm HMS}$ of the ten most massive structures correspond to the radii of the circles, each centre on the position of the corresponding sub-halo. The R$_{\rm HMS}$ of the BCG, in yellow, corresponds to the physical size of the yellow circle. The legend in the upper left panel of the plot shows the scaling size of the other sub-structures in the region, marked in white.  }
    \label{fig:wandering_all}
\end{figure*}

Finally, Fig. \ref{fig:wandering_all} displays the position of all the BHs in a projected stellar density map within 1 Mpc from the centre of the most massive halo in the \texttt{D9} simulation at redshift $z=0$ for
\texttt{REPOS} (top panel), \texttt{DYNMASS} (bottom left panel) and \texttt{DYNFRIC} (bottom right panel). We plot BHs with crosses of different colours depending on their distance $\Delta r$ from the closest sub-halo, comparing it to the $\rm R_{\rm HMS}$ of this sub-halo. Dark-blue crosses indicate BHs with $\Delta r <  \rm R_{\rm HMS}$, green crosses are for $\rm R_{\rm HMS} \leq \Delta r < 2 \times \rm R_{\rm HMS}$, and cyan crosses mark the wandering BHs, i.e. those with $\Delta r \geq 2 \times \rm R_{\rm HMS}$. We also plot the circles for the ten most massive sub-halos in the region, each having radius that is suitably scaled according to the value of its $\rm R_{\rm HMS}$. Each circle is centred on the sub-halo position and has a radius proportional to its $\rm R_{\rm HMS}$. The radius of the yellow circle corresponds to the actual physical value of the R$_{\rm HMS}$ of the BCG. For the purpose of readability, we adopted a different scaling for marking the size of the other sub-halos, as indicated in the figure legend.

Figure \ref{fig:wandering_all} shows that the \texttt{DYNFRIC} simulation exhibits a less numerous population of wandering BHs compared to \texttt{DYNMASS} (8 WBHs against 13 within the main halo R$_{\rm HMS}$), mainly located close to multiple BH systems. Most of the \texttt{DYNMASS} wandering BHs occupy the central region of the halo. However, crucial differences arise between \texttt{REPOS} and the outcomes of the other simulations: the overall population of BHs using the repositioning scheme, significantly decreases both in the core and in the outskirt of the halo. Most of the sub-halos, identified by stellar density peaks or with circles for the more massive ones, lacks a central BH.  

\subsection{Merger events and multiple BH systems} \label{Nmerg}
\begin{figure}
    \centering
    \includegraphics[scale = 0.5]{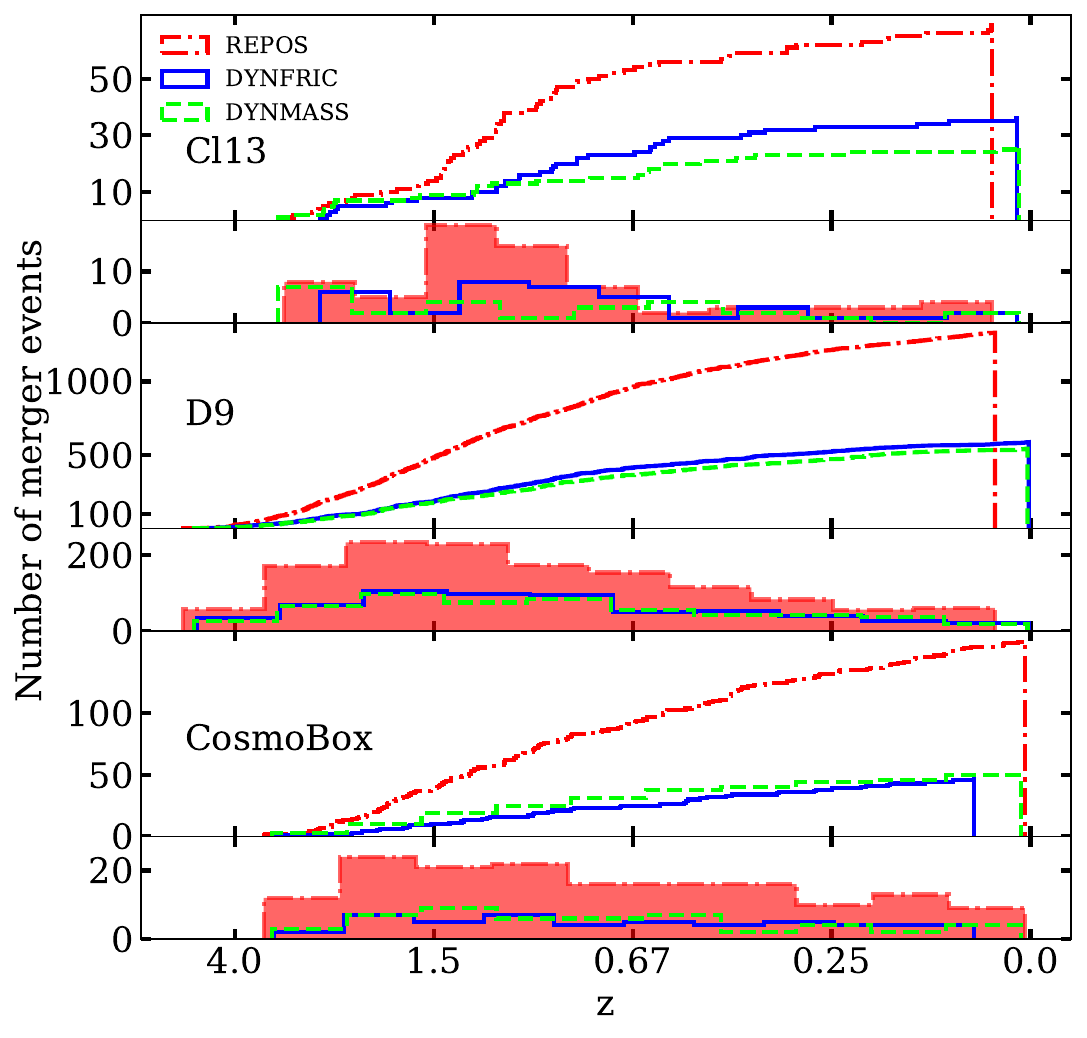}
    \caption{Cumulative and differential distributions of the number of merger events as a function of redshift. \texttt{Cl13}, \texttt{D9,} and \texttt{CosmoBox} are in the first, second, and third rows, respectively. The \texttt{REPOS} simulation results are marked with a dash-dotted red line (with a shaded area marking the differential distributions), the \texttt{DYNFRIC} simulations with a blue solid line and the \texttt{DYNMASS} simulations with a dashed green line. The curves stop at the redshift corresponding to the occurrence of last merger event.}
    \label{fig:nmerg}
\end{figure}
The occurrence and timing of merger events are highly sensitive to the prescription to control the BH dynamics. We observe that the \texttt{REPOS} simulations facilitate mergers when two BHs approach a distance $d_{\rm merg}$ (see Table \ref{tab:BHdetails}), making them merge on shorter timescales than when the other prescriptions are adopted (see Sect. \ref{zoom}).

Figure \ref{fig:nmerg} displays the cumulative (upper part) and the differential (lower part) distributions of the number of merger events as a function of the redshift, for the three simulations. \texttt{DYNFRIC} and \texttt{DYNMASS} simulations predict similar results, with \texttt{DYNFRIC} showing slightly more mergers in denser regions (\texttt{Cl13} and \texttt{D9}) and less mergers in the \texttt{CosmoBox} simulation. By contrast, all the simulations based on \texttt{REPOS} feature a significantly higher number of mergers, more than twice the number of mergers of both \texttt{DYNFRIC} and \texttt{DYNMASS} in the zoom-in simulations and more than three times in the \texttt{CosmoBox} simulation. Interestingly, the density peak of mergers for the repositioning model occurs at $z\simeq 1.5$ in all the simulations.

\begin{figure}
    \centering
    \includegraphics[scale = 0.43]{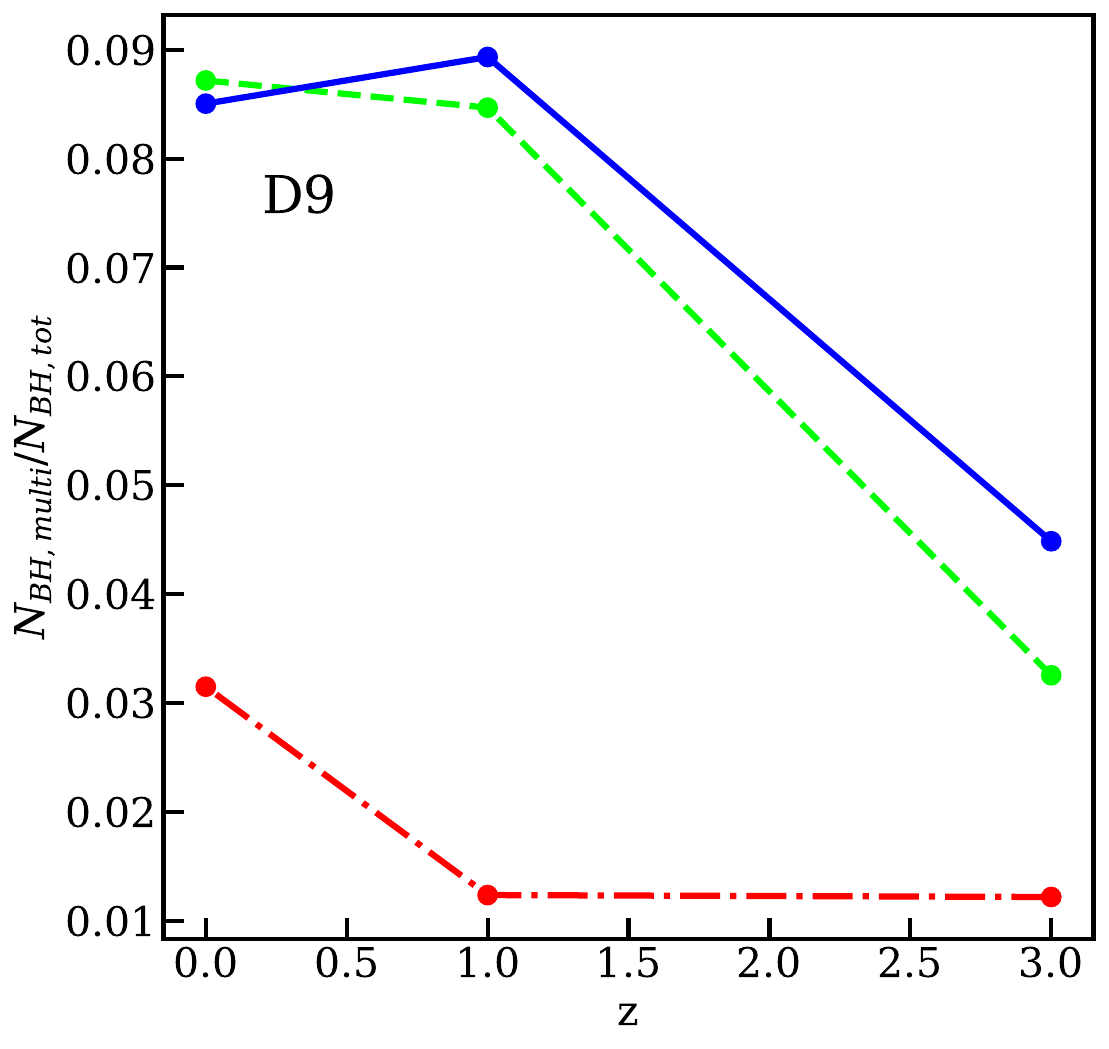}
    \caption{Ratio between the number of sub-halos hosting more than one BH and the total number of sub-halos hosting at least one BH in \texttt{D9} simulations. We consider \texttt{DYNMASS} (dashed green line), \texttt{DYNFRIC} (solid blue line) and \texttt{REPOS} (dot-dashed red line) at $z=3,1,0$. }
    \label{fig:multibhs}
\end{figure}
Before the merging events occur, one would expect to find structures hosting systems of two or more BHs. However, the increase of mergers in the repositioning scheme is not counterbalanced by an increase of multiple systems, i.e. of structures containing two or more BHs.

Figure \ref{fig:multibhs} shows the ratio between the number of sub-halos hosting more than one BH within the  $\rm R_{\rm HMS}$ to total number of sub-halos hosting at least one BH for $z=3,1,0$ in the \texttt{D9} simulations.
While the percentage of sub-halos hosting multiple BHs reaches more than the 8\% in \texttt{DYNFRIC} (blue) and \texttt{DYNMASS} (green) simulation, the \texttt{REPOS} shows a consistently lower percentage. The concurrence of a larger number of mergers and the nearly absence of multiple systems found for the \texttt{REPOS} simulations can be ascribed to extremely fast mergers, with multiple BHs coexisting within the same halo only for a rather short time. 

We note that the number of merger events is influenced both by the adopted model to trace the BH dynamics and by the specific seeding prescription adopted. In principle, a higher number of merging events may be associated with a more frequent seeding. This situation may arise when, following a merging event between BHs but not between their corresponding halos, one of the two halos remains orphan of its BH, while still matching the seeding conditions (see Table \ref{tab:BHdetails}). In this case, a new BH would be seeded at the centre of the orphan halo, thereby possibly contributing to increase the BH-BH merger rate. To quantify how this ‘seeding bias' affects the merging predictions of from each model, we report in Table \ref{tab:bhseeds} the number of seeded BHs in each simulation.

\begin{table}
\caption{Number of seeded BHs in each simulation for every sub-resolution prescription adopted.}
\label{tab:bhseeds}
\centering
\begin{tabular}{c c c c}
\hline 
\toprule
     Simulation & \texttt{REPOS} & \texttt{DYNFRIC} & \texttt{DYNMASS}  \\
\hline
\hline
     \texttt{Cl13}& 206 & 191 & 184 \\
     \texttt{D9}& 2239 & 2206 & 2169 \\
     \texttt{CosmoBox}& 328 & 314 & 323 \\
\hline
\hline
\end{tabular}
\end{table}

We observe that the increase in the total number of seeded BHs in the \texttt{REPOS} simulations is only marginal. The absence of a proportional rise in the number of seeding black holes despite the increase in mergers in the \texttt{REPOS} simulations can be addressed on the concurrence of two particular seeding conditions.

On the one hand, the seeding occurs exclusively within the main halos identified by the FoF halo finder, and not within the sub-halos identified by \texttt{SubFind}. On the other hand, BHs are not seeded in FoF groups which already contain a BH particle. Instead, the more frequent scenario in the \texttt{REPOS} simulations is the merging of two BHs initially belonging to two sub-structures contained within the same FoF halo. In this case, no further seeding takes place within the sub-halo which eventually remain orphan of its BH. 

Moreover, the simulations adopting the dynamical mass implementation predict less mergers in the \texttt{D9} and \texttt{Cl13} simulations (Fig. \ref{fig:nmerg}) and a slightly higher fraction of multiple BH systems at z=0 (Fig. \ref{fig:multibhs}) compared to \texttt{DYNFRIC}, even if the latter exhibits a higher number of seeded BHs (Table \ref{tab:bhseeds}). This circumstances already suggest that the dynamical mass model fails to reproduce some merger events, something that will be explored in detail in Sect. \ref{zoom}.

\subsection{The $M_{\rm *}$--$M_{\rm BH}$ relation} \label{magorrian}
\begin{figure*}
    \centering
    \includegraphics[scale = 0.47]{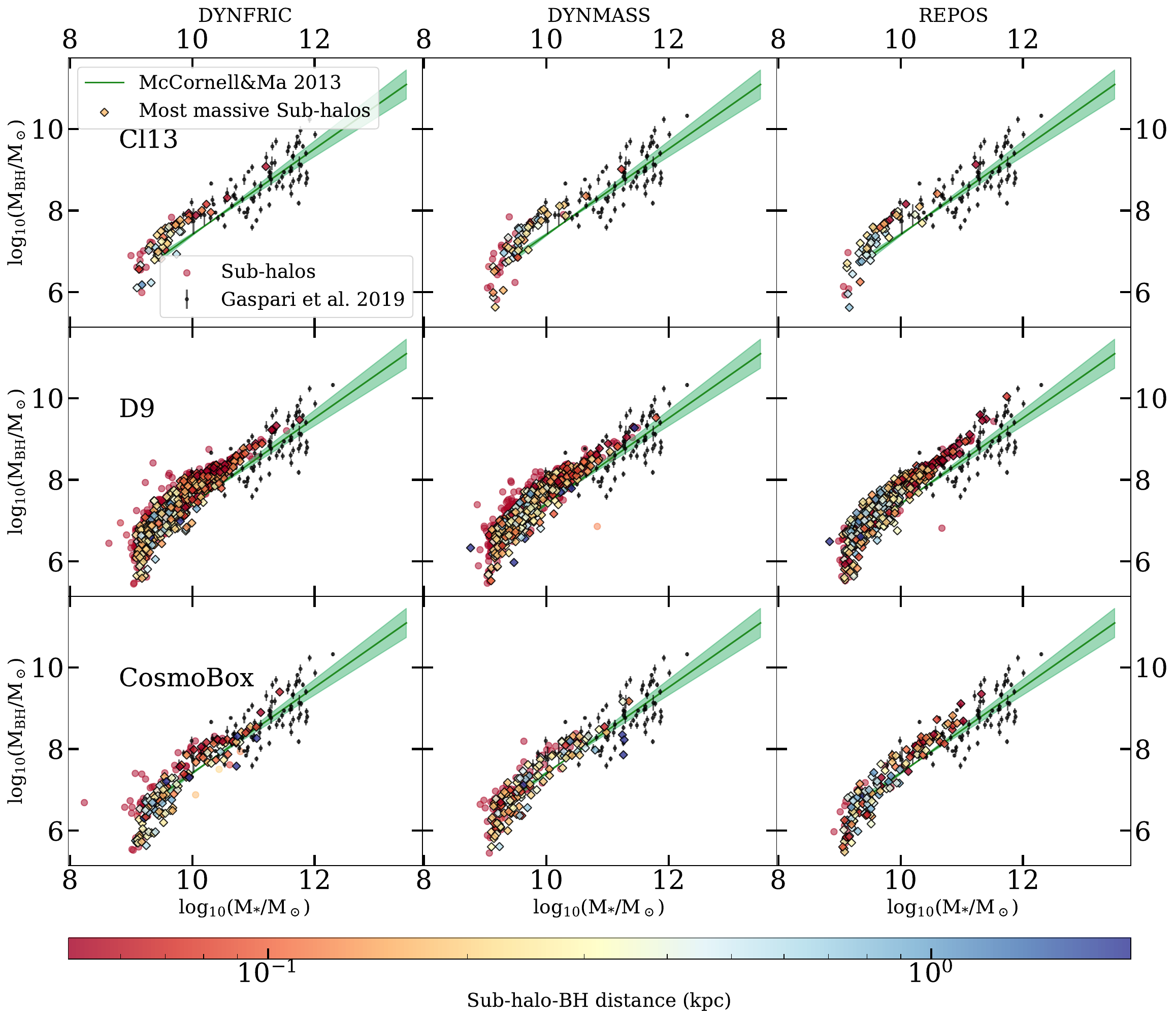}
  \caption{Relationship between BH mass and stellar mass of the host galaxies in the \texttt{Cl13} (upper panels), \texttt{D9} (central panels), and \texttt{CBox} simulations (lower panels). Each column contains the results obtained with a different BH dynamics prescription: \texttt{DYNFRIC} (left), \texttt{DYNMASS} (central), \texttt{REPOS} (right). The diamonds refer to the BHs associated with the main halos while circles correspond to sub-halos belonging to halos more massive than $10^{10}$ M$_\odot$.  Diamonds and circles are colour-coded according to the distance between the sub-halo centres and the associated BHs. For comparison, we also plot observational data from \protect\cite{gaspari_x-ray_2019} with black dots with errorbars, and the relation obtained by \protect\cite{mcconnell_revisiting_2013} with a green and shaded area. }
   \label{fig:magorrian}
\end{figure*}

A first diagnostic to investigate how the processes of star formation and galaxy evolution are intertwined with the evolution of the population of SMBHs is to look at the relationship between SMBH masses and stellar masses of the bulges of the host galaxies, the so-called Kormendy-Magorrian relation \citep{kormendy1993nearest, magorrian_demography_1998}. Figure \ref{fig:magorrian} shows this relation obtained by varying the implementation for the BH dynamics in all the simulations for the three initial conditions considered in our analysis.
In detail, for each halo with a mass larger than a threshold value $M_{\rm thr} = 10^{10} M_\odot$, we distinguish between the central galaxy, to be identified with BCG or Brightest Group Galaxy (BGG; marked with diamond symbols in Fig. \ref{fig:magorrian}), and the satellite galaxies, which are hosted within the substructures identified by {\tt{SubFind}} (marked with circles).
Since the \texttt{SubFind} algorithm does not split the diffuse stellar component of the Intra-Cluster light to the one bound to the BCG, at least for large-mass halos we adopted a fixed physical aperture within which measuring the BCG/BGG stellar masses  (see  for instance \cite{Marini_2020}). A fixed aperture is also commonly adopted for a fair comparison with observational measurements of BCG/BGG stellar masses \citep{Bassini_2019, Ragone_Figueroa_2018}). 
For the BCGs, we calculate the stellar mass of the stars belonging to the BCG/BGG within 70 kpc from the centre. As for the satellite galaxies, their stellar mass is computed by considering all the star particles that {\tt{SubFind}} assigns to the substructure. For the central galaxies, we associate the most massive BH within 70 kpc from the centre of the structure. For the satellite galaxies, we link the closest BH within their $\rm R_{\rm HMS}$. Symbols are colour-coded according to the distance between the galaxy centres and the associated BHs.
For reference, in each panel we also show the relation obtained by \cite{mcconnell_revisiting_2013} and the data for BCGs and BGGs as measured by \cite{gaspari_x-ray_2019}. We note in general that all the simulations reproduce rather well the observational relation, possibly with a slightly larger normalisation.
Besides proving the good calibration of the AGN feedback parameters, this agreement also demonstrates that this scaling relation is relatively insensitive to the details of the model adopted to account for unresolved DF on BH dynamics. 
\begin{table}
    \caption{Stellar mass, associated BH mass,  and its distance from the halo centre  for the most massive halo identified within each simulation.}
    \label{tab:pos_magorrian}
    \centering
    \makebox[\linewidth]{
    \begin{tabular}{cccc}
        \hline
        \toprule
         &$\rm log_{\rm 10}{(M_{\rm *}/M_\odot)}$ & $\rm log_{\rm 10}{(M_{\rm BH}/M_\odot)}$ & Distance \\
         & & & (kpc)\\
         \hline
         \hline
         \texttt{\textbf{Cl13}}& & & \\
         \texttt{REPOS} & 11.28 & 9.13 & 0.03 \\
         \texttt{DYNMASS} & 11.27 & 9.01 & 0.10 \\
         \texttt{DYNFRIC} & 11.25 & 9.07 & 0.07 \\
         \hline
         \texttt{\textbf{D9}}& & & \\
         \texttt{REPOS} & 11.75 & 10.0 & 0.10 \\
         \texttt{DYNMASS} & 11.89 & 9.52 &  0.12\\
         \texttt{DYNFRIC} & 11.83 & 9.47 & 0.07 \\
         \hline
         \texttt{\textbf{CosmoBox}}& & & \\
         \texttt{REPOS} & 11.36 & 9.35 & $ 1.25 \times 10^{-3}$ \\
         \texttt{DYNMASS} & 11.39 & 9.16 & 0.16 \\
         \texttt{DYNFRIC} & 11.47 & 9.4 & 0.08 \\
         \hline
         \hline
     \end{tabular}}
\end{table}
Table \ref{tab:pos_magorrian} presents data on stellar mass, associated BH mass, and the distance of the BH from the halo centre for the most massive halo within each simulation. The \texttt{DYNFRIC} simulations predict better centred BHs compared to the \texttt{DYNMASS} simulations, although all the distance values are below the resolution limit of the simulations.

As for \texttt{Cl13}, we note that the \texttt{DYNFRIC}, \texttt{REPOS} and \texttt{DYNMASS} simulations all produce very similar results.
 In all the three cases, the stellar mass of the BGG, and the mass of the hosted SMBHs are quite similar at $z=0$. 

In the \texttt{D9} region, the larger statistics of sub-halos helps to better understand what happens in different scenarios. Still, the \texttt{DYNFRIC} and the \texttt{DYNMASS} simulations produce comparable results, with the \texttt{DYNFRIC} simulation even further reducing the offset of the most massive BHs from the centre of their host galaxies. 
The results for the \texttt{D9} region using \texttt{REPOS} are in good agreement with the observational data for  $M_{\rm *}<10^{11}$ M$_\odot$. However, this simulation predicts BHs that are more massive compared to the other implementations, again due to excess of merging episodes, as already discussed. We note that the BH having the highest mass in the \texttt{D9} region results from the merging between the  BH \texttt{3} and the massive wandering BH \texttt{4} in Fig. \ref{fig:wbh_events} that reached the centre of the BCG from $z=0.06$ to $z=0$.

\begin{figure}
    \centering 
    \includegraphics[scale = 0.55]{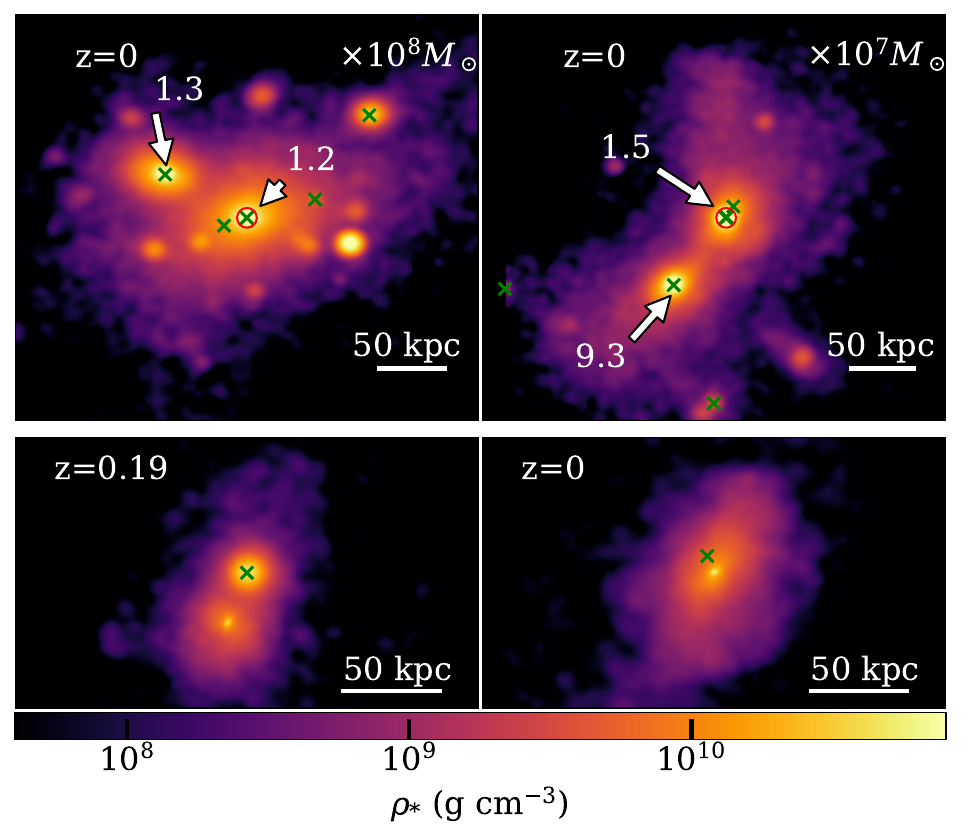}
    \caption{ Stellar density maps along the $z$-axis with depth of $350$ kpc (upper row) and $260$ kpc (bottom  row) of the massive sub-halos having a large separation between sub-halo centre and the associated BH in the \texttt{CosmoBox} simulation, using the {\tt DYNFRIC} model (blue diamonds in the corresponding plot in Fig.\ref{fig:magorrian}, for $\text{log}_{10}(M_{\rm *}/M_\odot) > 10$), at redshift $z=0$. BHs are marked by green crosses. Upper panels: two close encounters between structures. The halo centre is identified as a red circle. The arrows link to the values of the BH masses (in units as specified in the label). In both the left and right panels, the most massive BH belongs to the off-centre substructure. 
    Bottom panels: a merger event between two substructures at redshift $z=0.19$ (left panel) and the resulting merged halo at $z=0$ (right panel). The plot on the right side captures an off-centre BH while sinking toward the halo centre.}

    \label{fig:cosmo_disp}
\end{figure}

Finally, the \texttt{CosmoBox} results confirm the substantial agreement between our predictions and observations. The DF model further demonstrates its increased efficiency in centring the BHs compared to \texttt{DYNMASS}. 
Nonetheless, both simulations reveal three massive sub-halos with a stellar mass $> 10^{10.5}$ M$_\odot$, whose BHs have a significant displacement. The visual representation of these three occurrences is depicted in Fig. \ref{fig:cosmo_disp}, which shows the projection of the stellar density centred around these pathological sub-halos. In the first row, two analogous situations are presented: a close encounter between two substructures hosting BHs of different masses. The displacement obtained in Fig. \ref{fig:magorrian} follows from the association of the most massive BH to the central halo, marked in Fig.\ref{fig:cosmo_disp} with a red circle. 
This feature, rather than being a real off-centring of the BH, is a consequence of the method that we adopted to associate a BH to a given structure.

The third sub-halo, shown in the bottom right panel of Fig.\ref{fig:cosmo_disp}, hosts a truly off-centre BH. This is the outcome of a merger event between two substructures which took place at redshift $z=0.19$ (see bottom left panel). Since then, the BH belonging to the merging substructure did not have the time to sink to the centre of the merged sub-halos, hence the displacement at $z=0$ between the centre identified by \texttt{SubFind} and the BH.
\section{Analysis of individual events}\label{zoom}
Besides the analysis of the properties that emerge from  the population of BHs in each simulation, we perform a detailed study of the effect of different sub-resolution prescriptions on the dynamics of single BHs.

To better understand the BH dynamics, due to the different prescriptions to follow it, we adopt the following approach.
We freeze the configuration of BHs and of their host galaxies at two snapshots from the \texttt{Cl13-DYNMASS} simulation at $z=3$,  and $z=1.26$.  Using these snapshots as initial conditions, we run simulations using: the {\tt DYNFRIC} model introduced in Sect.\ref{newmodel}, the {\tt DYNMASS} method based on boosting dynamical mass at seeding (see Sect. \ref{dynamicalmass}), the {\tt REPOS}  method based on relocating the BH on the local potential minimum (see \ref{pinning}), and a fourth simulation not correcting to the BH position is introduced (\texttt{NOCORR} in the following). 
 The simulations having as initial conditions the snapshot at $z=3$ evolve until $z=2$, while the ones starting at $z=1.3$ reach $z=0.95$.
In this way, we trace the histories of BHs that, at the beginning of the simulations have the same position, mass and velocity, and are located in substructures with the same characteristics.
 Therefore, any difference between their subsequent orbits is purely driven by the different tracing methods adopted.

To ensure that the results were reproducible and marginally affected by the possible chaotic nature of a simulation, we carried out each of them twice, obtaining results which show marginal differences in the timings, while leaving the general results qualitatively unchanged.  
Clearly, restarting \texttt{DYNFRIC} and \texttt{REPOS} simulations from an output produced at a given intermediate redshift by the \texttt{DYNMASS} run drives to a structure evolution and dynamics that are different from those produced by using \texttt{DYNFRIC} and \texttt{REPOS} since the initial redshift, as described in the previous sections.
In particular, our analysis focuses on the evolution of binary BH systems, typically characterised by a massive BH located at the centre of a substructure, indicated as $\mathrm{BH_{\rm cen}}$, and a second displaced BH, the ‘satellite', labelled as $\mathrm{BH_{\rm sat}}$. 
We focus in the following on three events which represent three very different values of the central-to-satellite mass ratio between the BHs, $f_{\rm m} = \mathrm{m_{\rm BH,{cen}}}/\mathrm{m_{\rm BH,{sat}}}$; they are labelled as Event 1, Event 2 and Event 3 in the following. 
Event 1 consists of a binary system of two BHs initially displaced by $\mathrm{4}$ kpc from each other. The initial BH mass ratio is $ f_{\rm m} = 1.1 $, and both the BHs have a mass smaller than $m_{\rm DM}$ (see Table \ref{tab:zoom_tab}). For that reason, both masses are boosted in the \texttt{DYNMASS} simulation.
Event 2 involves a binary system of two BHs, initially at a distance of 1.5 kpc, with a higher mass ratio ($f_{\rm m} = 50$) compared to Event 1, but at the same redshift. Still, the two masses are both increased in the \texttt{DYNMASS} run. 
Event 3 consists of a binary system of two BHs initially separated by $10$ kpc at $\mathrm{z=1.26}$. The BHs have a  large mass ratio $f_{\rm m} = 313$. This time, only $\rm BH_{\rm sat}$, whose mass is less than $m_{\rm DM}$, undergoes the dynamical mass correction in the \texttt{DYNMASS} simulation.

\begin{table}[]
    \centering
    \caption{Initial characteristics of merger events described in Sect.\ref{zoom}.    }

    \label{tab:zoom_tab}
\begin{tabular}{p{1.07cm }p{1.4cm} p{1.4cm} p{1.5cm} p{1.2cm} }
    \hline
    \toprule
     & $m_{\rm BH,{cen}}$ (M$_{\odot}$) &  $m_{\rm BH,{sat}}$ (M$_{\odot}$) & $M_{sub}$ (M$_{\odot}$) & $\Delta r$ (kpc) \\
     \hline
     \hline
     Event 1 &  $3.23 \times 10^{5}$ & $3.63 \times 10^{5}$ & $6.96 \times 10^{10}$  & 4.08 \\
     Event 2 &  $3.60 \times 10^{6}$ & $7.20 \times 10^{5}$ & $8.11 \times 10^{10}$ & 1.48 \\
     Event 3 &  $6.39 \times 10^{8}$ & $2.04 \times 10^{6}$ & $4.50 \times 10^{11}$ & 10 \\
     \hline
     \hline
\end{tabular} 
\tablefoot{Column 1 and 2: initial true masses of the central ($m_{BH,{\rm cen}}$) and satellite ($m_{BH,{\rm sat}}$) BHs in the events described in Sect.\ref{zoom}, at the initial redshifts $z=3$ (for Events 1 and 2) and $z=1.26$ (for Event 3). In the Events 1 and 2, the BH masses of both $\rm BH_{\rm cen}$ and $\rm BH_{\rm sat}$ are boosted in the \texttt{DYNMASS} runs. On the other hand, we only boost the satellite BH mass in the Event 3. Column 3: mass of the sub-halo hosting the events. Column 4: initial distance between the two BHs.}
\end{table}
\begin{figure*}
    \centering
    \includegraphics[scale = 1.03]{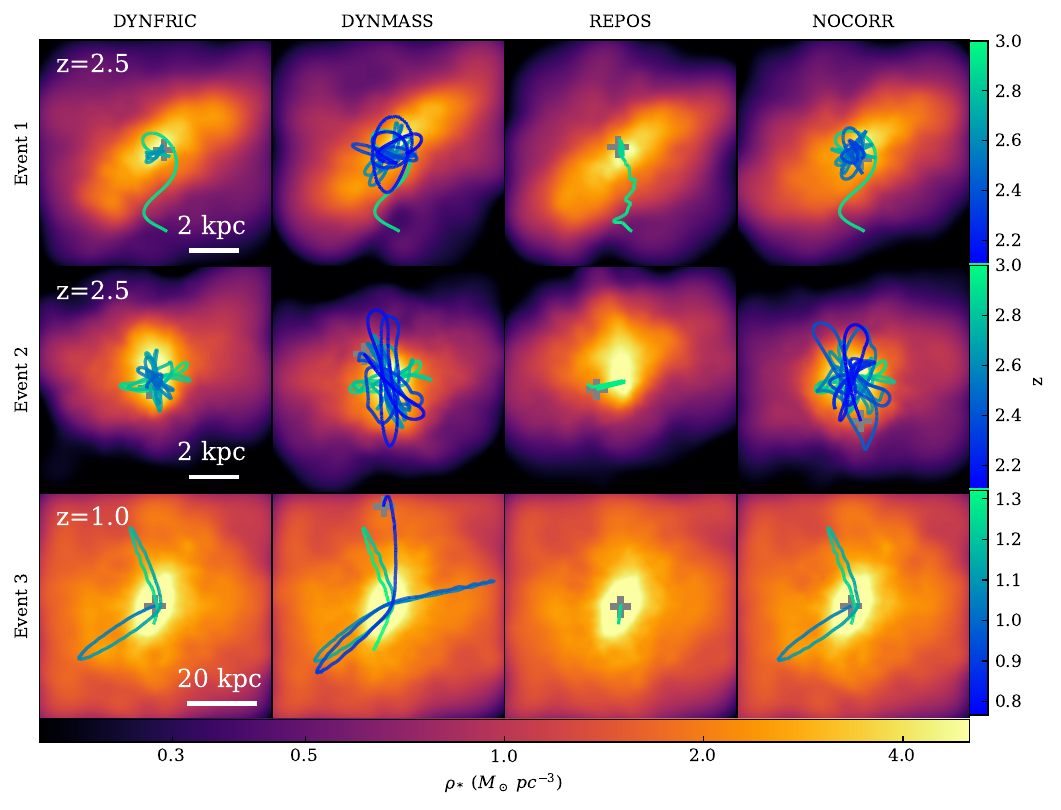}
    \caption{Stellar density map along the $z$-axis with depth of $10$ kpc in the top and central row and $70$ kpc in the bottom row, centred on the central BH ($\rm BH_{\rm cen}$) of Event 1 in the top row, Event 2 in the central row and Event 3 in the bottom row. The maps of stellar density refer to the redshift indicated on the left panel of each row. In each panel, the curve show the orbit of the satellite BH ($\rm BH_{\rm sat}$) involved in the merger event, colour-coded according to the redshift. The crosses indicate the last position of the satellite BH before either the merging event or the end of the simulations (see Sect. \ref{zoom}). The columns display the BH evolution using different sub-resolution prescriptions, from the left: using DF (first column), using a boosted dynamical mass (second), adopting the repositioning scheme (third) and finally without any correction for BH dynamics (fourth).}  
    \label{fig:all_mergers}
\end{figure*}
The characteristics of these three events are summarised in Table \ref{tab:zoom_tab}.
Figure \ref{fig:all_mergers} provides a graphical representation of the stellar density map at the redshift reported in the legend. The figure also includes the trajectory of the $\rm BH_{\rm sat}$, colour-coded according to redshift, within the substructure hosting that event. Each row displays a single event, and results from the \texttt{DYNFRIC}, \texttt{DYNMASS}, \texttt{REPOS} and \texttt{NOCORR} simulations are presented from left to right in each column.

Figure \ref{fig:single_plots} displays on the left column, for the different events in each row, the distance between $\mathrm{BH_{\rm cen}}$ and $\mathrm{BH_{\rm sat}}$, namely $\Delta r$, during the event. The dashed green line, the solid blue line, the dash-dotted red line and the short-dashed orange line indicate the \texttt{DYNMASS}, \texttt{DYNFRIC}, \texttt{REPOS} and \texttt{NOCORR} runs, respectively. The horizontal grey solid line represents the distance threshold which is necessary for the merger event $d_{\rm merg}$ to happen (see Table \ref{tab:BHdetails}).
The right side of Fig.\ref{fig:single_plots} focuses on the results of the \texttt{DYNFRIC} runs. In particular, we plot $\Delta r$ in the top panel, and the ratio between the DF force and the gravitational force (including the contribution of the DF correction), both for $ \mathrm{BH_{\rm cen}}$ (light-blue)  and  $\mathrm{BH_{\rm sat}}$ (light green) in the bottom panel. 
In the following subsections, we study each event separately. 
\subsection{Event 1}
\begin{figure*}
    \centering
    \includegraphics[scale = 0.45]{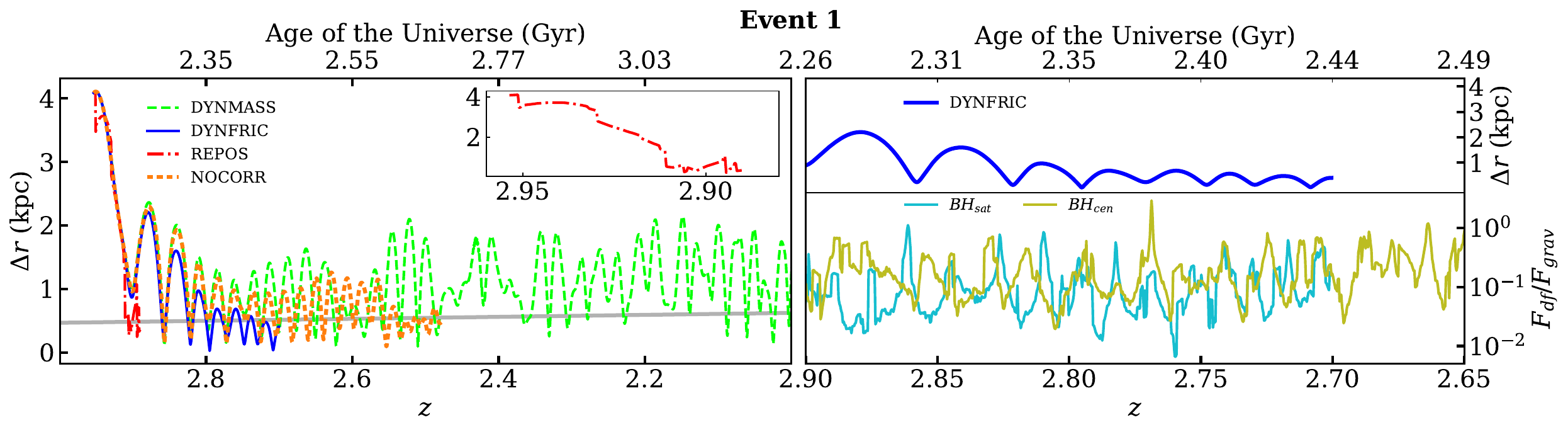}
    \includegraphics[scale = 0.45]{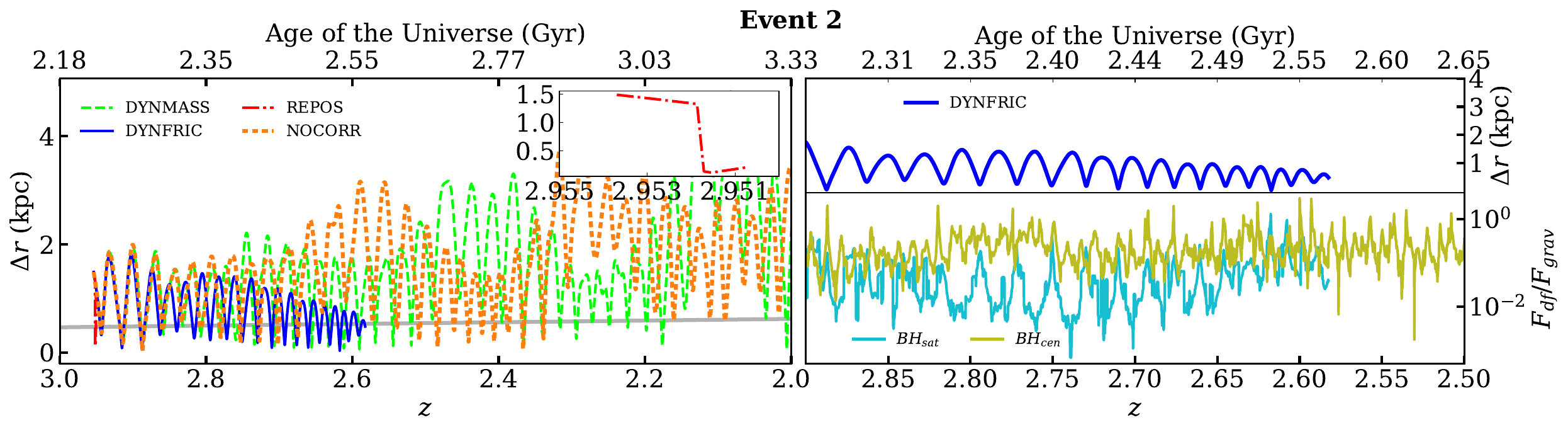}
     \includegraphics[scale = 0.45]{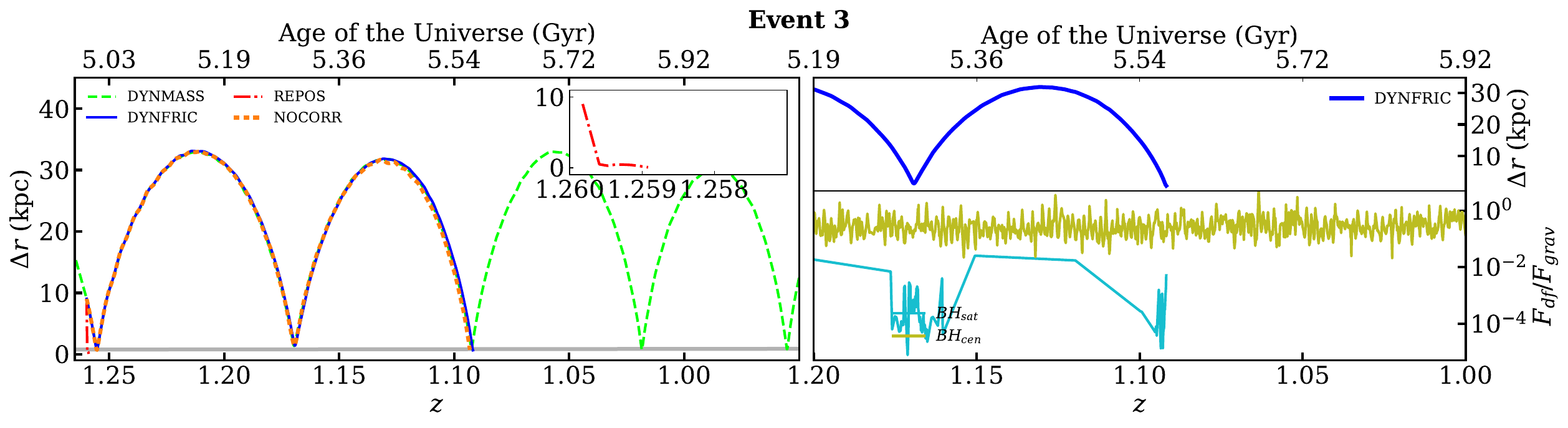}
    \caption{Left column: evolution of the distance between the central and the satellite BH in Event 1 (top), Event 2 (central), Event 3 (bottom). The dashed green line displays the evolution using the dynamical mass scheme ({\tt{DYNMASS}}), the blue solid line is for the DF scheme ({\tt{DYNFRIC}}), the dash-dotted red line for the repositioning ({\tt{REPOS}}), and the densely dashed orange line is for the case without any sub-resolution prescription ({\tt{NOCORR}}). The horizontal grey solid line represents the distance threshold which is necessary for the merger event $d_{\rm merg}$ to happen. Right column: evolution of the distance between the BHs for the {\tt{DYNFRIC}} case (top panel) and ratio between the DF and gravitational forces during each event.  }
    \label{fig:single_plots}
\end{figure*}
Looking at the upper row of Fig. \ref{fig:single_plots}, the simulation adopting the {\tt DYNFRIC} model exhibits oscillations with gradually decreasing amplitude and gently drives the two BHs toward the merger.
In the \texttt{DYNMASS} simulation, instead, the distance between the BHs exhibits persistent oscillations, which do not decrease in amplitude. Rather than driving the BHs to form a close pair, the enhanced dynamical mass intensifies their mutual gravitational attraction, causing collisions resulting in sustained relative distance fluctuations.
The simulation that does not employ any correction shows, after an initial gradual decrease of the distance, a second phase during which the two BHs keep oscillating with respect to each other, with a nearly constant amplitude. The merger between the two BHs, which occurs right after $z\sim 2.5$, is preceded by a sudden decrease in distance. The inset in the upper right side zooms on the results of the \texttt{REPOS} simulation. We observe a merger event occurring very rapidly, with the distance between the two BHs decreasing through discrete ‘jumps' as large as almost 1 kpc.

The figure shows that the two BHs can be closer than  $d_{\rm merg}$, yet without merging. We remind that, this represents a necessary, but not sufficient condition for the merger to happen. Indeed, merging requires the BHs to fulfill all the conditions listed in Table \ref{tab:BHdetails}. Whenever they are closer than $d_{\rm merg}$, they merge if their relative velocity is less than half of their surrounding sound speed and if the gravitational binding criterion defined in \eqref{mergercond_phi} is also satisfied.
From the visual representation of Fig.\ref{fig:all_mergers} we can infer that  \texttt{DYNFRIC}, \texttt{DYNMASS} and \texttt{NOCORR} show similar paths for $\rm BH_{\rm sat}$ until it first crosses the denser region of the sub-halo. Then, \texttt{DYNFRIC} and \texttt{NOCORR} simulations bound $\rm BH_{\rm sat}$ in the core of the host. Besides, \texttt{DYNMASS} reproduces constant oscillation at rather fixed apocentric distances. 
A detailed analysis of the reason why the merger does not occur in the {\tt DYNMASS} simulation shows that the merger failure is due to the enhanced relative velocity and to the difference of the gravitational potential between $\rm BH_{\rm sat}$ and $\rm BH_{ \rm cen}$. While crossing the distance threshold imposed by the merging criteria at the pericentre, the higher relative velocity and the larger potential difference compared to \texttt{DYNFRIC} prevent both merging conditions from being satisfied. The sound speed surrounding the BHs remains comparable in the two simulations. These considerations are still valid to explain why the \texttt{DYNMASS} simulation fails to reproduce the merger in Event 2  and Event 3.

On the other hand, the path of $\rm BH_{\rm sat}$ in the \texttt{REPOS} simulation is discontinuous and short compared to the other: through successive repositioning steps, the BH drops at the centre of the sub-halo.
The right panel of the plot highlights the features arising from the application of the DF correction in the {\tt DYNFRIC} model. In this case, the DF force responds to a local increase of stellar density. The ratio between the gravitational and the DF forces oscillates for both BHs with oscillation of comparable amplitude, indicating that they are orbiting around a central denser zone. Indeed, looking at the upper row of Fig. \ref{fig:all_mergers} we observe that the final merger will take place in the central, denser region on the substructure. 

\subsection{Event 2} \label{event2}
The second rows of Fig. \ref{fig:all_mergers} and of Fig.\ref{fig:single_plots} show the evolution of the distances during the Event 2.  Again, throughout the event, \texttt{REPOS} and \texttt{DYNFRIC} lead to immediate and gradual BH coalescences of the BHs, respectively. 

\texttt{DYNMASS} and \texttt{NOCORR} do not produce any merger, thus leaving $\rm BH_{\rm sat}$ swing around the central one. Thus, neither an ad hoc increase of the BH mass nor the N-body gravity solver without any correction is sufficient to dump the oscillations of the smaller BH.
Furthermore, the visual representation in Fig.\ref{fig:all_mergers} demonstrate that \texttt{DYNFRIC}, \texttt{DYNMASS} and \texttt{NOCORR} simulations all produce an initial tilt of the orbit of $\rm BH_{\rm sat}$ in the direction of the elongated denser region of the sub-halo. Then, in the \texttt{DYNFRIC} simulation, the contribution of DF leads the BH toward the centre. 
The right panel in Fig.\ref{fig:single_plots} illustrates the $F_{\rm DF}/F_{\rm grav}$ trend, showing that for this Event 2 the peaks of the contribution of the DF on $\rm BH_{\rm sat}$ usually take place in correspondence of the minimum  distance between the two BHs: as $\rm BH_{\rm sat}$ approaches $\rm BH_{\rm cen}$, it crosses the central region where the density of stellar particles is higher, thus leading to an increase of the DF acting on it. On the other hand, the DF acting on $\rm BH_{\rm cen}$ is relatively stronger during all the duration of the event, consistent with the more central position that it occupies.

\subsection{Event 3}
\label{event3}
Finally, the bottom rows of Fig.  \ref{fig:all_mergers} and of Fig. \ref{fig:single_plots} show the evolution during the Event 3.
In this case, at the lower redshift reached by restarting the simulations at $z=1.26$ to $z=0.96$, the mass of $\mathrm{BH_{\rm cen}}$ is more than 300 times higher than that of $\mathrm{BH_{\rm sat}}$. The latter is initially revolving around the central BH with large oscillations of approximately 30 kpc in amplitude. Once again, \texttt{REPOS} allows the two BHs, initially displaced by $\mathrm{10}$ kpc, to merge nearly instantaneously, with a ‘long-range teleporting' of $\rm BH_{\rm sat}$ from the outskirt to the centre of the substructure, reported in the top-right inset in the bottom left panel of Fig. \ref{fig:all_mergers}.
Numerical details of the repositioning scheme can affect the amplitude of the jump: its size could be smaller should different or additional constraints be taken into account (e.g. \citealt{Ragone_Figueroa_2018}, as also discussed in Sect. \ref{pinning}).

Distance oscillation in  \texttt{DYNFRIC}, \texttt{DYNMASS}, and \texttt{NOCORR} nearly coincide with each other. However, while \texttt{DYNFRIC} and \texttt{NOCORR} enable the merger to occur after a short time, this does not happen for \texttt{DYNMASS}. The reason for this lies in the different ways in which these simulations match the merging conditions: the distance threshold criterion is satisfied by all three simulations, but the other merging conditions remains unsatisfied for the \texttt{DYNMASS} simulation.  $BH _{\rm sat}$, having its dynamical mass boosted, keeps revolving around $\mathrm{BH_{\rm cen}}$ performing wide oscillations of 30 kpc in amplitude. In any case, the similarity of the oscillations described by \texttt{DYNFRIC} and \texttt{NOCORR} suggests that the DF correction in the former should be subdominant in this case.
The orbits of $\rm BH_{\rm sat}$ further demonstrate that \texttt{DYNFRIC}, \texttt{DYNMASS} and \texttt{NOCORR} produce almost identical results until $z\simeq 1.1$. In the \texttt{DYNMASS} simulation, the subsequent oscillations of the BH seem to be results of ‘kicks' that $\rm BH_{\rm sat}$ receives while crossing the denser core of the sub-halo.
In the right panel, we note that $F_{\rm df}/F_{\rm grav}$ for $\rm BH_{\rm sat}$ oscillates in phase with the distance between the BHs. We verified that this feature is driven by  wide oscillations of the gravitational force, which increases when approaching the galactic core where $\mathrm{BH_{\rm cen}}$ resides. Furthermore, the value of
$F_{\rm df}/F_{\rm grav}$ for $\rm BH_{\rm cen}$ is significantly higher than for $\rm BH_{\rm sat}$. This happens both because of the higher mass of the central BH and due to the strong DF correction due to the concentration of stellar particles in the central sub-halo region.

\vspace{0.3truecm}
In summary, these three events highlight the importance of adding a correction to the gravitational force onto BHs contributed by the unresolved DF, as demonstrated in Sect. \ref{zoom}. Furthermore, the description of such examples of merging events also reveals the significant limitations in correctly describing the dynamics of a system of two BHs of both the {\tt REPOS} (in the worst-case scenario considered here) and the {\tt DYNMASS} schemes. Indeed, our simulations based on \texttt{REPOS} generate ‘jumps' of the BHs, while \texttt{DYNMASS}, by artificially boosting the BH mass, produces two-body scattering effects, preventing the formation of a bound system.
Lastly, we point out that using an additional constraint on the relative velocity between the BH and the selected neighbor particles in the \texttt{REPOS} model (see Sect.\ref{pinning}) does not sensibly change  merging and sinking timescales when the two BHs are located within the same sub-halo, as in Event 1 and Event 2.

\section{Conclusions}
We introduce a novel approach to partially correcting for the effect of unresolved dynamical friction (DF) on the orbits of black hole (BH) particles in cosmological hydrodynamical simulations. We implement this correction using the \OG code. The main motivation of this study is the need for a more physically motivated method for preventing BH particles in simulations from spuriously leaving the host galaxies as a result of finite numerical resolution. The second aim of this study is to improve the description of BH-BH mergers. The DF correction implemented here has been extensively tested both in zoomed-in and fully cosmological simulations.
Specifically, we ran two zoomed-in simulations, one for a group-sized halo and another for a cluster-sized halo, and a cosmological volume with a box size of $(16  \ \mathrm{comoving \ Mpc})^3$. 

We assessed the performance of the  new model by carrying out each simulation with other prescriptions for the BH dynamics: the repositioning scheme and the adoption of a large dynamical mass to reinforce the unresolved DF contribution. To distinguish the effects of variations in the prescription of BHs dynamics, all simulations were run with identical resolution and sub-resolution physics. We summarise the main conclusions of our analysis as follows: 
\begin{itemize}

    \item \textbf{Offset}: The simulations using the continuous repositioning of BH particles on the local potential minimum ({\tt REPOS} scheme) exhibit the smallest offset between the BH and the centre of the host sub-halo. Our model to correct unresolved DF ({\tt DYNFRIC} scheme) provides an accurate centring, even outperforming, at $\mathrm{z<1,}$ the scheme based on boosting the dynamical mass of BH particles at seeding ({\tt DYNMASS} scheme). See Sect.\ref{centring} and Fig.\ref{fig:distancestot}.

    \item\textbf{Wandering BHs}: The {\tt DYNFRIC} simulations produce the smallest population of wandering BHs, which are predominantly found in multiple BH systems. The {\tt REPOS} prescription features the most numerous and massive set of wandering BHs, as close encounters between galaxies and large-scale potential environments favour the spurious repositioning of BHs outside their host sub-halos. See Sect.\ref{wandering}, Fig.\ref{fig:wbh}, Fig.\ref{fig:wbh_events}, and Fig.\ref{fig:wandering_all}.

    \item\textbf{Mergers}: The overestimate of merger events in the {\tt REPOS} simulations in denser environment leads to an excess of galaxies deprived of a central BH, an effect that is more pronounced in the denser environment of a massive galaxy cluster. See Sect. \ref{mergers}, Fig. \ref{fig:nmerg}, and Fig. \ref{fig:wandering_all}.

    \item\textbf{$\mathbf{M_{\rm \textbf{*}} -M_{\rm \textbf{BH}}}$ relation}: The good agreement with the observational data, that is, the ability  of all the simulations
 to reproduce the observed $M_{\rm *} -M_{\rm BH}$ scaling relation, demonstrates a relative insensitivity of this diagnostic to the particular prescription adopted, while it is sensitive to the choice of the parameters regulating BH accretion and the ensuing AGN feedback efficiency. See Sect. \ref{magorrian} and Fig. \ref{fig:magorrian}.
\end{itemize}

To delve deeper into the details of how different prescriptions affect the orbits of BH particles, we focused on specific BH--BH interaction events. To disentangle the possible diversity in the substructure evolution between different simulations, we restarted the simulations from two snapshots of the group-sized halo at $z=3$ and at $z=1.26$ to explore how BHs respond to different methodologies governing their dynamics in the same environment. In this analysis, we also carried out simulations without any sub-resolution prescription to correct the orbits of BH particles.
The results of this further analysis, introduced in Sect. \ref{zoom},  Fig. \ref{fig:all_mergers}, and Fig. \ref{fig:single_plots}, can be summarised as follows:
\begin{itemize}
    \item \textbf{DF}: Our novel model, adding the DF correction, predicts damped orbits of the satellite BHs, which gradually approach the centre of a host galaxy, and eventually form a close BH--BH pair. 
    \item \textbf{Dynamical mass}: The large BH dynamical masses in the {\tt DYNMASS} scheme can lead to spuriously strong interactions between BHs, which could delay or even prevent merger events. Moreover, such interactions eventually inhibit the BHs from satisfying the gravitational boundedness criteria for merging.
    The overall results, in terms of the quality of the description of orbital decay in BH pairs, can be worse than those obtained without any prescription to correct BH dynamics for unresolved DF. 
    \item \textbf{Repositioning}: Our implementation of the {\tt REPOS} method allows extremely rapid mergers preceded by large and sudden movements of the satellite BH, which promptly reaches the central region of the sub-halo with a few `jumps'; these jumps can eventually span several kiloparsecs.
\end{itemize}

In summary, an extensive analysis of the BH population arising in different simulations demonstrates that our novel implementation of the correction for unresolved DF force acting on BH particles introduced in this paper provides a robust and reliable description of the DF exerted on BHs by their surroundings. This model achieves at least the same performance as other ad hoc numerical prescriptions in terms of centring BHs in their host halos, while significantly reducing the population of wandering BHs, and overcoming the limitations of this prescription in terms of its ability to describe BH-BH merger events.
Thanks to the recent work by \cite{sala2023supermassive}, the \OG code is now equipped with a spin-evolution model that, coupled to the DF correction, is now supported by a more precise description of the BH dynamics. 
The compelling estimate of the sinking timescales of BHs onto host halos  (see Appendix \ref{appensixsinking}) arising from the application of the DF correction proposed in this work merits further investigation \citep[see][]{Taffoni_2003}, which we will present in an upcoming work (Damiano et al. in prep.). Such an analysis is crucial for further studies  of BH merger rates based on cosmological hydrodynamical simulations \citep[e.g.][]{degraf.et.al.2023}, and will serve as a powerful tool for fully exploiting the potential of gravitational wave astrophysics.
Furthermore, examining the interaction between BH dynamics and galaxy evolution may offer fresh insight. As discussed in the present study, adding a DF correction alters both the demography of the BH population and their merger timescales, which have an impact on where and how they interact with the surrounding gas through accretion and heating.

\begin{acknowledgements}
We are grateful to the referees for their thoughtful feedback, which has significantly improved our work.
We thank Gian Luigi Granato for a number of useful discussions and comments on this paper.
Simulations have been carried out at the CINECA Supercomputing Center (Bologna, Italy), with computing time assigned through ISCRA-B calls and through CINECA-INAF and CINECA-UNITS agreements, and at the computing center of INAF -- Astronomical Observatory of Trieste \citep{Bertocco.etal.2020,Taffoni.etal.2020}. We acknowledge the CINECA award under the ISCRA initiative, for the availability of high-performance computing resources and support.
This paper is supported by: the Italian Research Center on High Performance Computing Big Data and Quantum Computing (ICSC), project funded by European Union - NextGenerationEU - and National Recovery and Resilience Plan (NRRP) - Mission 4 Component 2, within the activities of Spoke 3, Astrophysics and Cosmos Observations; by the PRIN 2022 PNRR project (202259YAF) "Space-based cosmology with Euclid: the role of High-Performance Computing", and by a INAF Grant within the "Astrofisica Fondamentale" funding scheme. We acknowledge partial financial support from the INFN Indark Grant and by the INAF project "CONNECTIONS" (COllaboratioN oN codE development for future Cosmological
simulaTIONS).  Klaus Dolag acknowledges support by the COMPLEX project from the European Research Council (ERC) under the European Union’s Horizon 2020 research and 
innovation programme grant agreement ERC-2019-AdG 882679 as well as support by the Deutsche Forschungsgemeinschaft (DFG, German Research Foundation)
under Germany’s Excellence Strategy - EXC-2094 - 390783311 
We acknowledge the European Union's HORIZON-MSCA-2021-SE-01 Research and Innovation programme under the Marie Sklodowska-Curie grant agreement number 101086388  - Project acronym: LACEGAL.
\end{acknowledgements}

\bibliographystyle{aa}
\bibliography{lib_paper} 

\begin{appendix} 

\section{Velocity distribution surrounding the BHs }
\label{appendix_distrbution}
One of the hypotheses on which our model to account for unresolved DF is based, is that the velocity distribution function in the vicinity of a BH particle can be expressed as in Eq. \eqref{velocitydistrib}.
\begin{equation} \label{eqdispvel}
   \rm  f(v) = \frac{3}{4 \pi \epsilon_{BH}^3} \sum_{i}^{N(<\epsilon_{BH})} \delta(v-v_{BH})
\end{equation}
meaning that the local velocity distribution is proportional to the PDF within the softening length. 
The possibility to recover, through this formula, the velocity distribution surrounding the BH strongly depends on density and resolution. 
To address the question of how many particles are necessary to accurately sample the background velocity distribution, we investigated whether we are able to recover the velocity distribution of the particles surrounding the BHs by considering only the particles within the BH softening length.
 We report here few examples of typical frequent conditions that we encountered.
 From snapshots of the \texttt{D9} and \texttt{Cl13} simulations, by visual inspection we identified BHs whose host galaxies were not undergoing mergers or tidal disruptions. For those cases, we verified that a Maxwellian distribution of all the stars belonging to such galaxies properly describes the velocity distribution.
Then we selected star particles within different distances from each BH, and compared the distribution of the relative velocities between star particles and BHs. We considered the following limiting distances: 30 kpc, 5 $\epsilon_{\rm BH}$, 2 $\epsilon_{\rm BH}$ and $\epsilon_{\rm BH}$.
Table \ref{tab: masses and stars} reports the BH mass and the stellar mass enclosed within 30 kpc from the halo centre for the examples we describe below.
\begin{figure}[!htb]
    \centering
    \includegraphics[width=1\linewidth]{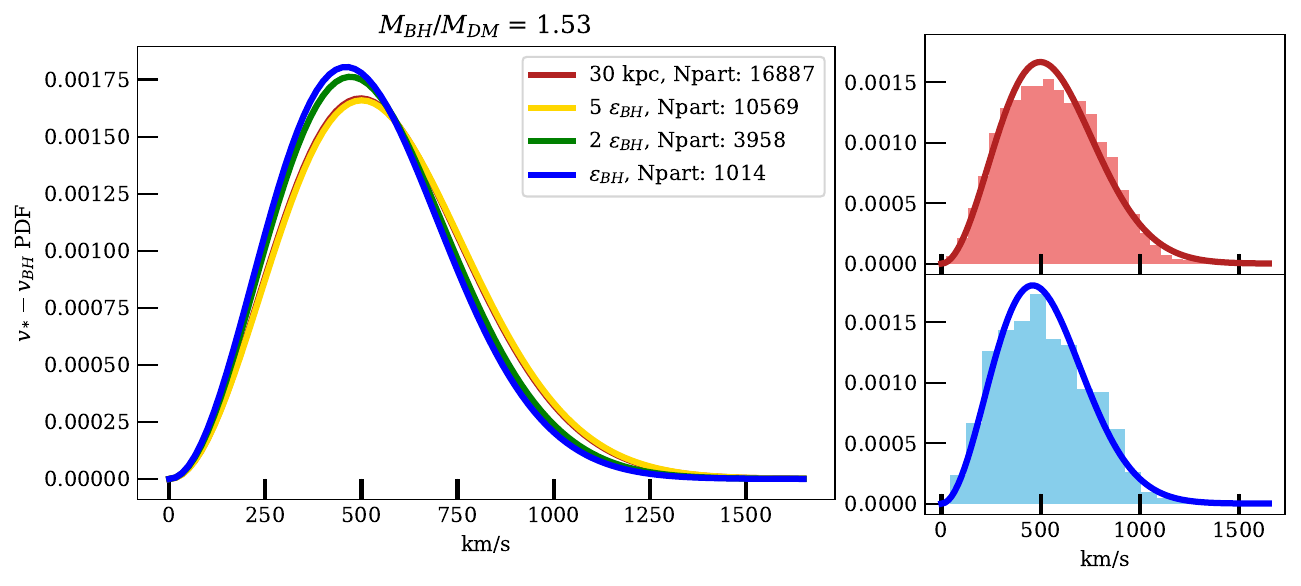}
    \caption{Probability distribution functions  of the relative velocity between a massive BH (see \ref{tab: masses and stars}) in the $\mathtt{Cl13}$ simulation using DF at redshift $z=2.5$,  and the surrounding stars within 30 kpc (enclosing all the stars in the sub-halo) (red line), 5 $\epsilon_{BH}$ (yellow line), 2 $\epsilon_{BH}$ (green line) and $\epsilon_{BH}$ (blue line). In the left panel, we compare the best fitting Maxwellian PDFs for these four cases. The right plots superimpose the PDF measured from the simulation with the Maxwellian fit for the 30 kpc case (top) and the $\epsilon_{BH}$ case (bottom).}
    \label{fig:fig_067}
\end{figure}
\begin{table}[]
    \centering
    \caption{Mass of BH and stellar mass of the host sub-halos within 30 kpc whose velocity distribution in shown in Fig. \ref{fig:fig_067}, \ref{fig:fig_010} and \ref{fig:fig_085}. }
    \label{tab: masses and stars}
\begin{tabular}{p{ 1.2cm }p{2cm} p{2.8cm}}
    \hline
    \toprule
     & $ \rm M_{\rm BH}$ (M$_{\odot}$) &  $ \rm M_{\rm *}$ (<30kpc) (M$_{\odot})$\\
     \hline
     \hline
     \ref{fig:fig_067} &  $7.14 \times 10^{7}$ & $3.44 \times 10^{10}$\\
          \ref{fig:fig_010} &  $3 \times 10^{6}$ & $2.14 \times 10^{9}$\\
               \ref{fig:fig_085} &  $2.57 \times 10^{8}$ & $6.14 \times 10^{10}$\\
     \hline
     \hline
\end{tabular} 
\end{table}

Figure \ref{fig:fig_067} compares the velocity probability distribution function (PDF) for a selected BH hosted at the centre of a galaxy having more than 16000 star particles. The dynamical BH mass corresponds to 1.53 times the DM particle mass.  
While less than 10\% of the star particles lies within the softening length (see legend), their velocity distribution can be quite accurately described by a Maxwellian distribution, as shown in the left plot.  Interestingly, the distributions that best fit in the surroundings of the BH ($2 \ \epsilon_{BH}, \epsilon_{BH} $) shift their peaks down to lower relative velocities. Particles with lower relative velocity compared to the BH contribute the most to the computation of the DF correction, according to Eq. \eqref{new}. 
\begin{figure}
    \centering
    \includegraphics[width=1\linewidth]{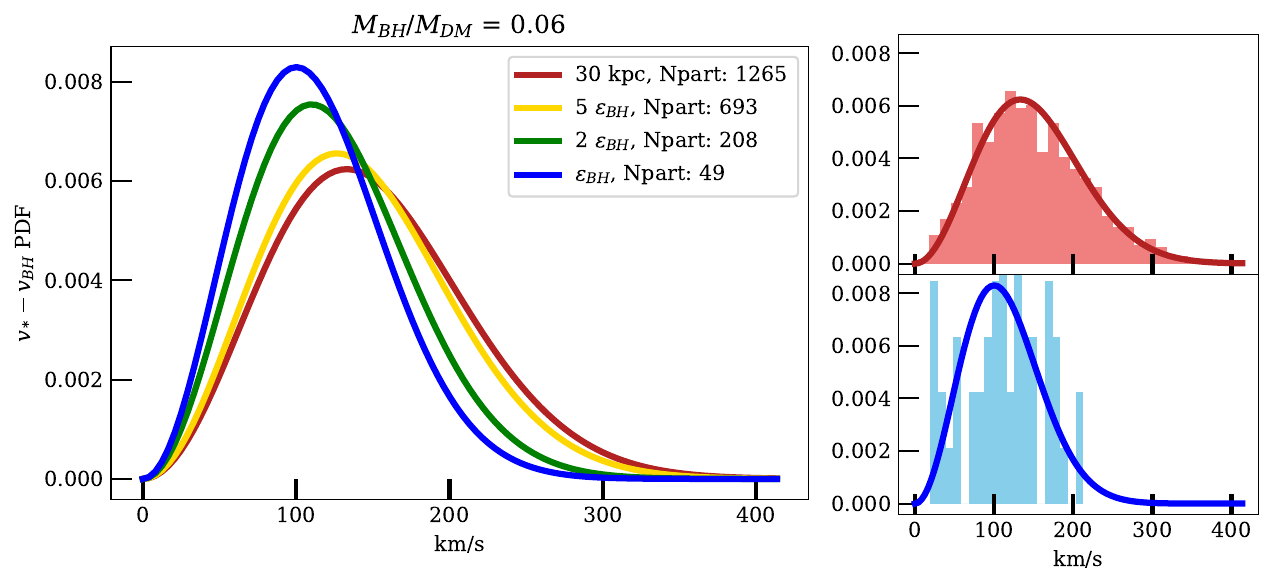}
    \caption{PDF fitting of the velocity distribution as Fig. \ref{fig:fig_067} but for a smaller BH at $z=2.5$ in the \texttt{Cl13} simulation.}
    \label{fig:fig_010}
\end{figure}
The same trend can be observed when the host sub-halo is resolved with fewer star particles.

For the case presented in Fig.\ref{fig:fig_010},  the  particle number within the entire galaxy is 1265 and only 49 of them are found within $\epsilon_{BH}$, thus contributing to our DF correction. Moreover, the BH mass reaches only 6\% of the mass of the DM particles. Even if the system is less resolved than that described in Fig. \ref{fig:fig_067}, the dominant contribution to the PDF of velocity in the vicinity of the BH comes from particles having a velocity relative to the BH smaller than 200 km/s, whose contribution is again dominant in the correction of the DF. 
Finally, when the BH local velocity distribution deviates from the host galaxy velocity distribution profile, this technique can trace the local velocity distribution.
Although the DF force is a {\em global} effect, whose ‘long-range' contribution is already captured by the N-body solver, the added correction should have a {\em local} nature.

\begin{figure}
   \centering
    \includegraphics[width=1\linewidth]{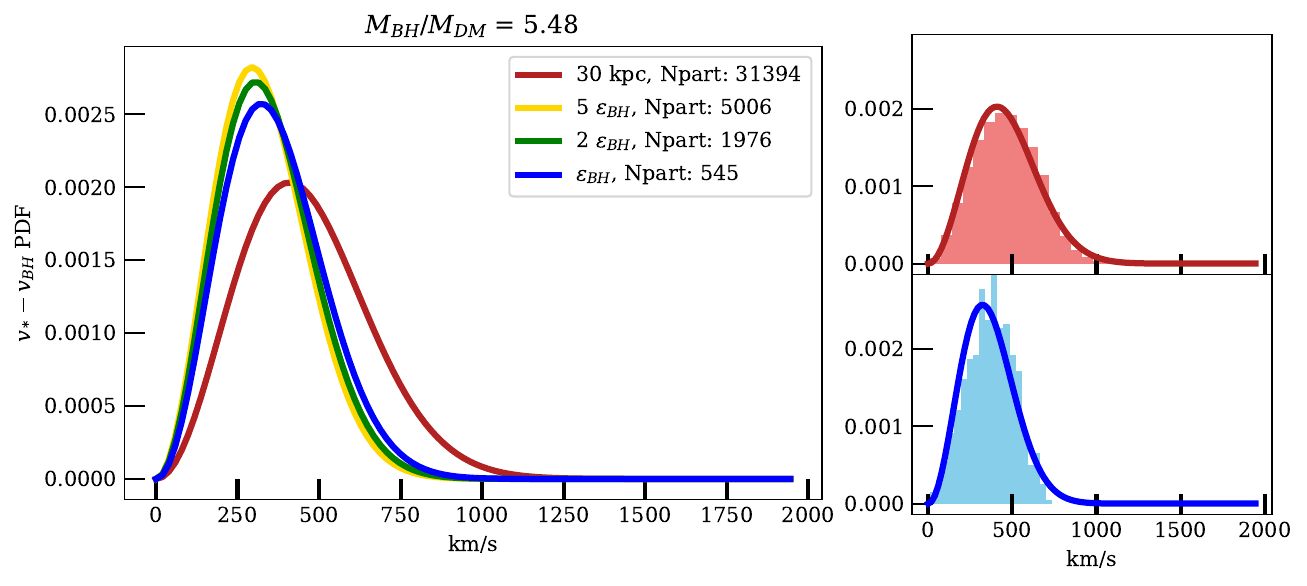}
    \caption{PDF fitting of the velocity distribution as in Fig. \ref{fig:fig_067} for a BH in the \texttt{D9} simulation using the DF correction at $z=2.5$.  The relative velocity PDF around the BH is locally different from the global distribution.}
    \label{fig:fig_085}
\end{figure}

For instance, the velocity distributions in Fig. \ref{fig:fig_085} describe a peculiar situation. As the BH is displaced from the sub-halo centre of 4 ckpc/h, the velocity distribution in the BH's surroundings differs from that of the sub-halo. Thus, enclosing only particles in the local vicinity of the BH leads to velocity distributions different from the one of the entire system. 

Figures \ref{fig:fig_010}, \ref{fig:fig_067}, and \ref{fig:fig_085} are only three examples that are representative of the vast majority of the cases we obtained when performing tests on the validity of Eq. \ref{eqdispvel}. 
Thus, we conclude from this analysis that for well-resolved systems the velocity PDF from particles within the softening length reproduces fairly well the PDF that characterises the BH underlying velocity field. Although for less-resolved systems the velocity dispersion is less constrained, the contribution from particles with low velocity relative to the BH is dominant. These particles are those that provide the dominant contribution in the DF correction of Eq. \ref{new}.

\section{Validating the DF correction in an isolated halo} \label{appensixsinking}
To validate the DF model introduced in this work, we test its performance in a controlled experiment in which a BH particle is placed on an initially radial orbit within an isolated DM halo, which follows a Navarro, Frenk and White (NFW, \citealp{nfw}) mass density profile. We then compute the sinking timescale of a BH falling inwards to the centre of the halo, when the halo is described at increasing resolution. In this idealised scenario, the mass density and velocity distribution of the halo are fully specified, thus allowing for a direct comparison with analytical results from the original DF formula from \cite{chandrasekhar}.
Furthermore, the limited computational cost of these simulations enables one to perform multiple simulations at different resolution. This allows us to validate the DF correction of Eq. \eqref{newmodel}, which depends in fact on the numerical resolution. 

We generate the initial conditions of a NFW DM-only halo using the \texttt{MAKEGALAXY} code \citep{springelwhite}
For the tests shown in this section, we adopted the same halo parameters of \cite{genina2024}. The halo has a virial radius of $\rm R_{vir}=350.54$ kpc, a virial mass of $\rm M_{vir}=10^{13}$~M$_\odot$ and concentration $\rm c = 4.38$. 
 Having verified that the halo density stabilses into the expected NFW mass density profile and the velocity distribution defines a Maxwellian distribution with dispersion $\rm \sigma(r)$,  we seed a BH at $20$ kpc from the centre with a mass of $ \rm 10^9$~M$_\odot$, setting it onto a circular orbit around the halo centre. We repeat this seeding process by progressively increasing mass and force resolution by sampling the halo with $\rm 10^{5}, 10^{6}, 10^{7}, 5 \cdot 10^{7}$ particles within $ \rm R_{vir}$. 
 We adopt Plummer-equivalent softenings derived from \cite{Powe2003}. Once seeded, the BH has the same softening length as the surrounding particles. We report in Table \ref{tab: halo_resolution} the number of particles, the mass of DM particles, and their softening for each resolution. Every simulation is carried out twice, with and without the DF correction introduced in Sect. \ref{newmodel}.
 
\begin{table}[htb!]
    \centering
    \caption{Numerical resolution of each simulated NFW halo. }
    \label{tab: halo_resolution}
\begin{tabular}{p{ 2cm }p{2cm} p{2.8cm}}
    \hline
    \toprule
     $ \rm N_{part}$ &  $ \rm M_{DM} $($ \rm M_\odot$) &$ \rm \epsilon_{DM}$ (kpc)\\
     \hline
     \hline
     $10^5$& $10^{8}$& 4.43\\
      $10^6$& $10^{7}$& 1.44\\
       $10^7$& $10^{6}$& 0.44 \\
        $5 \cdot 10^7$& $2 \cdot 10^{5}$& 0.20 \\
     \hline
     \hline
\end{tabular} 
\tablefoot{For each resolution, we report the number of particles (Column 1), the corresponding DM particle mass (Column 2) and its softening (Column 3), computed according to the prescription by \cite{Powe2003}.}
\end{table}

\begin{figure}
    \centering
    \includegraphics[width=1\linewidth]{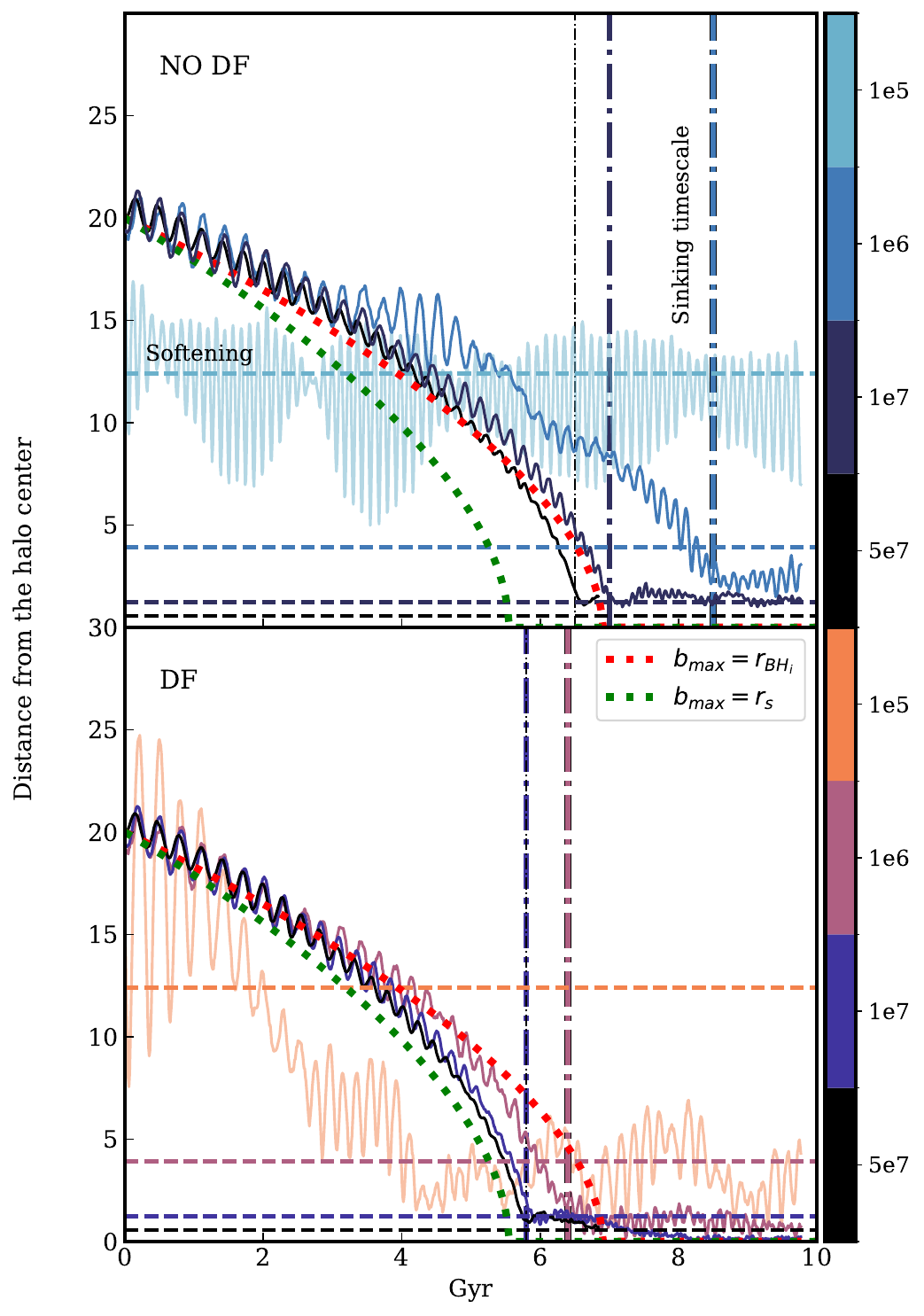}
    \caption{ Sinking timescales for a BH orbiting in a NFW DM halo. We plot the distance to the halo centre found through the shrinking sphere algorithm \citep{Powe2003} as a function of time. We plot the results from numerical simulation without any correction to the BH dynamics (top panel) and with the DF implementation introduced in Sect. \ref{newmodel} (bottom panel). The colourmap indicates the number of particles sampling the halo such that darker lines refer to higher numerical resolution. The dashed-dotted vertical lines are indicative sinking timescales and horizontal lines report the softening lengths, both are colour-coded with the simulation they refer to. The dotted lines are the analytical preditions for an infalling BH into a NFW halo, choosing the same BH and halo parameter of the numerical simulations, but varying the maximum impact paramter from $\rm b_{max}=20$ (red) and $\rm b_{max}= 80$ (green).} 
    \label{fig:sinking_halo}
\end{figure}
 Given this ideal setup, where the density profile is known and the velocity distribution is Maxwellian, we can compare the sinking timescale $ \rm t_s$ obtained from numerical simulation with the analytical expectation provided by Eq. \eqref{eq:DF_BT}. Following a similar procedure to that presented in \cite{rodriguez2018new}, it is possible to compute analytically the path of a BH initially seeded on a circular orbit at $ \rm \vec{r}_{{BH}_i}$ and with initial velocity $ \rm \vec{v}_{{BH}_i}$ assuming that its orbit remain circular. A detailed description of these calculations as well as a more extended version of the tests shown here below will be presented in Damiano et al. (in prep). 

Figure \ref{fig:sinking_halo} shows the distance of the BH from the centre of the halo, determined using the shrinking sphere algorithm \citep{Powe2003}, across all simulations performed. The upper and lower panels display the results without and with DF correction, respectively. In both panels, higher resolution corresponds to darker lines, as indicated by the corresponding colour maps. The softening lengths, obtained by multiplying the corresponding Plummer-equivalent values by a factor of 2.8, are marked by horizontal lines, colour-coded according to the corresponding resolution. The vertical dashed-dotted lines indicate approximate sinking timescales which correspond to the time at which the orbits stop further shrinking.
 The dotted green and red curves correspond to the analytical predictions of the BH decaying orbital radius, which are obtained by setting $\rm b_{min}=GM_{BH}/(v_{BH}^2+2/3 \sigma^2)$ \citep{just2011dynamical} for the minimum value of the impact parameters appearing in the Coulomb logarithm and choosing $\rm b_{max}=r_{{BH}_i} = 20$ kpc (red), corresponding to the initial BH orbit radius, and $\rm b_{max}=r_{s} = 80$ kpc (green), corresponding to the scale radius of the NFW halo.

In the test at the lowest resolution, applying the DF correction enables the BH to spiral inward toward the halo centre. Without any correction, the BH remains at a roughly constant distance near the softening length, showing persistent oscillations.
As the resolution increases, without the addition of any DF correction (NODF case), the sinking timescale $\rm t_s$ decreases from $\rm t_s = 8 \ Gyr$ when sampling the halo with $\rm 10^6$ particles, to $\rm t_s = 6.5 \ Gyr$ when using $ \rm 5 \cdot 10^7$ particles. On the other hand, the DF correction significantly shortens $\rm t_s$ compared to the NODF case, at fixed resolution. For example, when sampling the halo with $\rm 10^6$ particles, $ \rm t_s \simeq 6.5 \ Gyr$, which is only reached at the highest resolution in the NODF case. In addition, when using the DF correction, the higher-resolution simulations (using $10^7$ and $5 \cdot 10^7$ particles) converge toward similar sinking timescales. 
Comparing these results with the analytical predictions, we observe that without the DF correction, the highest-resolution simulation approaches the prediction based on the smaller value of the impact parameter. Using the DF correction, the simulations with a number of particles from $10^6$ to $5 \cdot 10^7$ show timescales that are fully consistent with the analytical ones. In conclusion, we validate our novel DF correction by setting idealised initial conditions for a halo with an NFW profile, adopting parameters similar to those used by \cite{genina2024}. These controlled experiments enable a reliable comparison with the analytical results derived from theoretical calculations. Our results show that applying the DF correction proposed in Sect.\ref{newmodel} drives to sinking timescales which are fully consistent with those predicted from the DF formula in Eq. \eqref{eq:DF_BT}.

\end{appendix}

\end{document}